\documentclass[a4paper,11pt]{article}
\pdfoutput=1
\usepackage{amssymb,amsmath,amsfonts,makeidx,placeins,pbox,multirow}
\usepackage{graphicx,rotate,subcaption,color,slashed,cite,caption,epstopdf,verbatim}
\usepackage{longtable,tabu}
\usepackage{array}
\usepackage{pstricks}
\usepackage{color}
\usepackage{amsmath}
\usepackage{mathtools}
\usepackage{amssymb}
\usepackage[normalem]{ulem}

\newcolumntype{P}[1]{>{\centering\arraybackslash}p{#1}}
\usepackage[colorlinks=true,
            linkcolor=magenta,
            urlcolor=blue,
            citecolor=blue]{hyperref}
\newcommand\Zdis{Z_{\rm dis}}
\newcommand\Zexc{Z_{\rm exc}}
\evensidemargin=\oddsidemargin
\newcommand{\beq}{\begin{equation}}
\newcommand{\eeq}{\end{equation}}
\newcommand{\bea}{\begin{eqnarray}}
\newcommand{\eea}{\end{eqnarray}}

\def\bar {\overline}

\def\to {\rightarrow}

\def\bea {\begin{eqnarray}}
\def\eea {\end{eqnarray}}
\def\n {\nonumber}

\def\barr{\begin{array}}
\def\earr{\end{array}}
\def\to{\end{rightarrow}}

\def\gev{\ensuremath{\mathrm{Ge\kern -0.1em V}}}

\usepackage{color}
\definecolor{amber}{rgb}{1,0.45,0}

\usepackage{tikz}
\usetikzlibrary{positioning,arrows}
\usetikzlibrary{decorations.pathmorphing}
\usetikzlibrary{decorations.markings}

\newcommand\sigmab{\Delta_b}
\begin{document}

\tikzset{
vector/.style={decorate, decoration={snake,amplitude=.6mm}, draw=red},
scalar/.style={dashed, draw=blue},
fermion/.style={draw=black, postaction={decorate},
        decoration={markings,mark=at position .55 with {\arrow[draw=black]{>}}}}
%fermion/.style={draw=black, decorate, decoration={markings,mark=at position .5 with {\arrow[draw=black,thick]{>}}}}
}
\begin{center}
{\Large \bf Alternative signatures of the quintuplet fermions at the LHC and \\
\vspace{0.3cm}
future linear colliders} \\
\vspace*{0.4cm} {\sf   $^{a}$Nilanjana Kumar\footnote{nilanjana.kumar@gmail.com},  $^{b}$Vandana Sahdev \footnote{vandanasahdev20@gmail.com}} \\
\vspace{6pt} {\small } {\em  $^{a,b}$University of Delhi, New Delhi, \\ $^{a}$CCSP, SGT University, Gurugram, Delhi-NCR, India} \\
\normalsize
\end{center}
%==========================================================================================================
\vspace{-0.4cm}
%===========================================================================================================
\begin{abstract}
Large fermionic multiplets appear in different extensions of the Standard Model (SM), which are 
essential to predict small neutrino masses, relic abundance of the dark matter (DM) and 
the measured value of muon anomalous magnetic moment (muon (g-2)). 
Models containing quintuplet of fermions ($\Sigma$), along with other scalar multiplets,  
can address recent anomalies in the flavor sector while satisfying the constraints 
from the electroweak physics. In standard scenarios,
the exotic fermions couple with the SM particles directly and 
there exists a strong limit on their masses from collider experiments such as the 
Large Hadron Collider (LHC). In this paper, we choose a particular 
scenario where the quintuplet fermions are heavier than the scalars, 
which is naturally motivated from the muon (g-2) data. 
A unique nature of these models is that they predict non-standard signatures 
at the colliders as the quintuplet fermions decay via the scalars once produced at the colliders. We study these non-standard interactions and provide alternative search strategies for these exotic 
fermions at the LHC and future linear colliders (such as $e^+e^-$ colliders). We also discuss their exclusion and discovery limits. For the 
doubly charged quintuplet fermion ($\Sigma^{\pm\pm}$), discovery is possible 
with 5$\sigma$ significance at integrated luminosity of 
3000 fb$^{-1}$ at 14 TeV LHC if $M_\Sigma\leq 980$ GeV. For 
the singly charged quintuplet fermion ($\Sigma^\pm$), the discovery is challenging 
at the LHC but there might be a possibility of 5 $\sigma$ discovery with 1000 fb$^{-1}$ 
luminosity at $e^+e^-$ collider for $M_\Sigma\leq 700$ GeV. 
\end{abstract}
%==========================================================================================================
\vspace{-0.4cm}
%===========================================================================================================
\tableofcontents
\vspace{-0.4cm}
%----------------------------------------------------------------------------------------------------------
\section{Introduction}
%----------------------------------------------------------------------------------------------------------
\label{sec:intro}
\vspace{-0.1cm}
The Standard Model of particle physics has been observed with great 
precision at the experiments. The last missing piece, the Higgs boson, was discovered 
by ATLAS \cite{ATLAS:2012yve} and CMS \cite{CMS:2012qbp} at the Large Hadron Collider (LHC) in 2012. However, it has some shortcomings -- it is not possible to account for 
the small neutrino masses or the existence of dark matter in the SM, for instance.
It is also difficult to explain the current measurement of muon (g-2) \cite{Muong-2:2021ojo} 
and the outcomes of flavor experiments \cite{Zyla:2020zbs} within the SM framework.
 
The small neutrino masses can be achieved 
by introducing exotic particles at a high scale via the effective
dimension-five Weinberg operator at tree level vis-a-vis the seesaw mechanism. 
These extra particles correspond to 
a heavy fermion singlet, a scalar triplet and a fermion triplet in type I, II and III seesaw mechanisms, 
respectively \cite{Mohapatra:1979ia,Konetschny:1977bn,Magg:1980ut,
Schechter:1980gr,Cheng:1980qt,Lazarides:1980nt,Mohapatra:1980yp,Foot:1988aq}.
However, there are other models, where the exotic particle content 
involves one or more larger multiplets of scalars and fermions 
together. These models \cite{Liao:2010cc,Babu:2009aq,Tavartkiladze:2001by,
Kumericki:2012bh,Mohapatra:1974hk,Senjanovic:1975rk,KumarAgarwalla:2018nrn,
Ko:2015uma,Cai:2011qr,Cirelli:2009uv} not only explain the small neutrino masses but 
also provide an explanation for the muon (g-2) data, flavor anomalies 
while also predicting a suitable dark matter (DM)
candidate in some cases. Such models also predict exotic signatures at the colliders.

A good example of such models are cascade seesaw-like 
scenarios
\cite{Liao:2010cc,Babu:2009aq,Tavartkiladze:2001by,Kumericki:2012bh}, where the neutrino mass is generated via a higher 
dimensional $(5+4n)$ operator, where $n=1$ 
is the minimal scenario with three generations of Majorana quintuplets $\Sigma_R$, with 
hypercharge $Y=0$ and a scalar quadruplet $\Phi$, with hypercharge $Y=-1$.
Another example is the left-right symmetric (LRS) framework \cite{Mohapatra:1974hk,Senjanovic:1975rk}
with an $SU(2)_R$ quintuplet where the gauge group is extended 
to $SU(3)_C\times SU(2)_L\times SU(2)_R\times U(1)_{B-L}$. 
In these models, the presence of a right-handed neutrino 
in the particle spectrum is essentially governed by the gauge 
structure and hence, naturally explains the origin of light 
neutrino masses. Further, the neutral component of the quintuplet 
can be a DM candidate~\cite{KumarAgarwalla:2018nrn,Ko:2015uma}. 
In models such as R$\nu$MDM \cite{Kumericki:2012bh, Cai:2011qr}, 
purely radiative neutrino masses are generated while 
also providing a viable DM candidate \cite{Cirelli:2009uv}.
Models with both exotic fermions and charged scalars can also be motivated from 
the little Higgs scenarios \cite{Low:2002ws}, where the global 
symmetry is $SU(6)/Sp(6)$. These models successfully explain the flavor anomalies \cite{Kumar:2020yco} and
the signatures can be studied at the colliders. Multicharged 
fermions also appear in models of warped extra dimensions and 
some other models as shown in Ref: \cite{Eichten:1983hw,Cabibbo:1983bk,Cirelli:2005uq,
delAguila:2008pw,Ma:2014zda,Chakrabarty:2020jro,delAguila:2010es,Wang:2017sxx}.

In this paper, we are motivated by a model with one quartet and one 
septet scalar fields and quintuplet Majorana fermions (3 copies) 
\cite{Nomura:2017abu}. The interactions between the SM $SU(2)_L$ 
lepton doublets and these large multiplets induce neutrino masses which 
are suppressed by small VEV's of the quartet and/or the septet and also by 
the inverse of the quintuplet fermion mass ($\sim$ TeV). 
As a result, the scale of the neutrino Yukawa coupling can be reached to 
less than $\mathcal{O}(1)$ \cite{Nomura:2017abu,Kumar:2019tat}. Further, this type of model is safe from any
quantum anomaly, given the zero $U(1)_Y$ charge of these quintuplet fermions. 
There is no contribution to $SU(2)$ gauge anomaly as well.

The contents of the quintuplet are doubly charged fermions, 
singly charged fermions and a neutral fermion\cite{Nomura:2017abu}.
Signatures of quintuplet fermions at LHC have been studied in Ref:\cite{Yu:2015pwa,Chen:2013xpa}.
These fermions are also good candidates for exotic particle search in 
the future collider experiments \cite{Ozansoy:2019kdq,Guo:2016hjt}.
In most of the phenomenological studies involving the quintuplet fermions, 
they decay directly into the SM particles. Even in scenarios as in 
Ref: \cite{Kumericki:2012bh}, where interactions 
between the quintuplet fermions and the scalar multiplets are allowed, 
the fermions can not decay into the scalars as the scalars are 
slightly heavier than the fermions. However, in our scenario\cite{Kumar:2019tat}, a small 
mass difference between the scalar multiplets  
and the fermionic quintuplet is naturally implied 
from muon $(g-2)$ and in such a way that the quintuplet fermions are heavier than the scalars.

In this paper, we study the signatures of the singly and the doubly charged 
quintuplet fermions ($\Sigma_R$) at the hadron collider ($pp$ collider) and linear colliders
such as $e^+e^-$ colliders, 
with each having its own advantages. Although the LHC has a much higher energy reach, 
the $e^+e^-$ colliders provide a cleaner environment for distinguishing the 
signal from the background \cite{CLICDetector:2013tfe} and are more suitable for precision 
measurements \cite{Lukic:2014mha}. With many $e^+e^-$ colliders, such as FCC-ee, ILC and CLIC in 
development stage, it would be interesting to study the discovery potential 
of the exotic quintuplet fermions at these colliders 
\footnote{This model also offers interesting 
signatures for the charged scalars which we do not address here.}.

Cases studied so far include multilepton and multijet 
signatures of the quintuplet fermions, both at the LHC \cite{Yu:2015pwa, Kumericki:2012bh}
and the linear colliders \cite{Zeng:2016tmw,Ozansoy:2019kdq,Guo:2016hjt}.
As already stated, the quintuplet fermions in the scenario considered here decay into the SM particles via the scalars which leads to final states containing a large number of 
leptons and jets. Such signals for quintuplet fermions have not been studied before. Even though it is difficult to reconstruct the masses of the fermion quintuplets or the scalars, given these many particle final states, we show that by carefully 
choosing the final states from amongst the many possibilities, it is indeed possible to reconstruct both the masses. The masses of the exotic fermions, 
such as vectorlike quarks and leptons, are constrained to be more than $\sim$ 1 TeV
\cite{Rappoccio:2018qxp, Buckley:2020wzk} and $\geq$ 740 GeV \cite{CMS:2018cgi} respectively.
However, for this model, we choose to explore masses much below 1 TeV as well as masses 
larger than 1 TeV, considering the nonstandard decay modes 
of the quintuplet fermions.
For the singly charged scalar, the direct search limit from LEP is 80 GeV \cite{ALEPH:2013htx}. 
However, the limit on the production cross-section of the singly charged scalar
as a function of its mass is given in Ref:\cite{CMS:2019bfg}. 
Multilepton states where the doubly charged scalar has decayed into two 
same sign leptons are already studied in Ref:\cite{ATLAS:2017xqs,Chakrabarty:2020jro,ATLAS:2021jol}. 
The lower mass limit ranges from about 230-870 GeV. However, 
these studies are based on several assumptions which 
do not apply to our case. In the analysis done here, the mass of the scalars is chosen to be well below 
the mass of the quintuplet fermions to facilitate the decay of the quintuplet fermion via the scalars.

The rest of the paper is arranged as follows: The model is described in section \ref{sec:model}. 
Collider signatures at the LHC and linear collider are described in 
sections \ref{sec:pp} and \ref{sec:ee}, respectively. 
We discuss the results and conclude in section \ref{sec:outlook}.
%----------------------------------------------------------------------------------------
\section{Model}
\label{sec:model}
%-------------------------------------------------------------------------------------------
\vspace{-0.1cm}

We consider a simple scenario based on the models proposed in Ref:\cite{Nomura:2017abu,Kumar:2019tat}. 
In the model of Ref:\cite{Nomura:2017abu},
the interaction among the SM $SU(2)_L$ lepton doublets and scalar multiplets, quartet 
($\phi_4$, $Y=1/2$) and septet ($\phi_7$, $Y=1$), 
can induce neutrino masses, while preserving the $\rho$ parameter. In order to achieve that 
it is also necessary to introduce a quintuplet Majorana fermion $\Sigma_R$, with $Y=0$. 
The neutrino masses are suppressed by the small VEVs of the quartet or septet and an
inverse of the quintuplet fermion mass, which explains the smallness of the neutrino mass
while also relaxing the Yukawa hierarchies.
In order to make the generation of neutrino mass more natural, 
additional quintet scalar field ($\phi_5$, $Y=0$) with a
non-zero VEV can be introduced \cite{Kumar:2019tat}. As a result, tree level neutrino mass is forbidden and the
quartet scalar is an inert scalar while neutrino mass is generated at one-loop level.
The model is further constrainted from lepton flavor violation (LFV) and the muon anomalous
magnetic moment ($\Delta a_{\mu}$).

In these models, the interesting point to note is that the quintuplet fermion 
$\Sigma_R$ does not decay to standard model particles directly. Instead, it happens via 
its interaction with the charged scalars. Our objective is to study this particular scenario. Hence, 
we choose a simplistic scenario with quintuplet fermion, $\Sigma_R$ ($Y=0$) and the 
quartet scalar, $\phi_4$ ($Y=1/2$), along with their components \footnote {Even in the models with 
more than one scalar multiplet, the components mix during mass diagonalization and 
charged and neutral scalar mass eigenstates are obtained.}

\vspace{-0.6cm}

\begin{align}
& \Phi_4 = \left( \varphi^{++}, \varphi^{+}_2, \varphi^{0}, \varphi^{-}_1 \right)^T, \nonumber \\
& \Sigma_R = \left[ \Sigma_1^{++}, \Sigma_1^{+}, \Sigma^{0}, \Sigma_2^{+}, \Sigma_2^{++} \right]_R^T,
\end{align}

($\Sigma_1^{\pm}$, $\Sigma_1^{\pm \pm}$) and ($\Sigma_2^{\pm}$,$\Sigma_2^{\pm \pm}$) 
are combined to make singly and doubly charged Dirac fermions, which we denote as $\Sigma^{\pm}$
and $\Sigma^{\pm \pm}$ respectively while $\Sigma^0_R$ 
remain a neutral Majorana fermion. The masses of each component are given by $M_\Sigma$ at
the tree level where mixing between the SM leptons is negligibly small.
The Yukawa interaction can be written as,  

\vspace{-0.6cm}

\begin{align}
-\mathcal{L}_{Y}
&=
(y_{\ell})_{ii} \bar L_{L_i} H e_{R_i} +(y_{\nu})_{ij} [\bar L_{L_i} \tilde\Phi_4 \Sigma_{R_j} ]
 +  (M_{R})_i [\bar \Sigma^c_{R_i} \Sigma_{R_i}] + {\rm h.c.},
 \end{align}

The components of the quintuplet fermion can decay into quartet scalars and SM
leptons via the Yukawa interaction given by the second term of the above equation as, 
%--------------------------------------------------
\begin{align}
-{\cal L}_{yuk}& \supset (y_{\nu})_{ij} [\bar L_{L_i} \tilde\Phi_4 \Sigma_{R_j} ]+{\rm h.c.} \n \\
&= (y_\nu)_{ij}
\left[
\bar\nu_{L_i} \left(\frac{1}{\sqrt2}\Sigma_{R_j}^0 \varphi^{0*} +\frac{\sqrt3}{2} \Sigma_{1R_j}^+ \varphi_1^- +  \frac{1}{2}\Sigma_{2R_j}^+ \varphi_{2}^- +  \Sigma_{1R_j}^{++} \varphi^{--}  \right)\right.
\n\\
&\left. +
\bar\ell_{L_i} \left(\frac{1}{\sqrt2}\Sigma_{R_j}^0 \varphi_1^-  + \frac12 \Sigma_{1R_j}^+ \varphi^{--} +  \frac{\sqrt3}{2}\Sigma_{2R_j}^- \varphi^{0*} +  \Sigma_{2R_j}^{--} \varphi_{2}^{+}  \right)\right] +{\rm h.c.}
\n\\
\label{eq:yukawa}
\end{align}
%----------------------------------------------------
The coupling $y_{\nu}$, as given in Eq.~\ref{eq:yukawa}, can be constrained 
from observables such as $\nu$-mass, $\Delta a_{\mu}$ and flavor observables. 
Following the benchmark points in \cite{Nomura:2017abu, Kumar:2019tat}, this 
coupling varies between (0.001--1), depending on the scalar particle content 
of the model. As we are studying a more general scenario, for the phenomenological purpose, 
we choose a very conservative limit of 0.1 for $y_{\nu}$. 

The components of the quintuplet can be produced via gauge interactions given by,
\begin{align}
\bar \Sigma_R \gamma^\mu i D_\mu \Sigma_R \supset  &  \bar \Sigma^{++} \gamma^\mu \left( 2 e A_\mu + 2 g c_W  Z_\mu  \right) \Sigma^{++} + 
\bar \Sigma^{+} \gamma^\mu \left( e A_\mu + g c_W  Z_\mu  \right) \Sigma^{+} \nonumber \\
&  - \sqrt{2} g   \bar \Sigma^{++} \gamma^\mu W_\mu^+  \Sigma^{+} - \sqrt{3} g   \bar \Sigma^{+} \gamma^\mu W_\mu^+ \Sigma_R^{0} - \frac{\sqrt{5} g}{\sqrt{2}}   \bar \Sigma^{+} \gamma^\mu W_\mu^+ \Sigma^{0c}_R \nonumber \\
& - \sqrt{2} g   \bar \Sigma^{+} \gamma^\mu W_\mu^-  \Sigma^{++} - \sqrt{3} g   \bar \Sigma^{0}_R \gamma^\mu W_\mu^- \Sigma^{+} - \frac{\sqrt{5} g}{\sqrt{2}}   \bar \Sigma^{0 c}_R \gamma^\mu W_\mu^- \Sigma^{+},
\label{eq:gauge_sigma}
\end{align}
where $s_W(c_W) = \sin \theta_W (\cos \theta_W)$ with the Weinberg angle $\theta_W$\footnote{The components of $\phi_4$ can also be produced at the collider via the gauge interaction, 
as shown in Ref:\cite{Nomura:2017abu}.}.

The relevant gauge
interactions associated with $\phi_4$ can be obtained from the following kinetic term
\begin{equation}
|D_\mu \Phi_4|^2 \supset \sqrt{\frac{3}{2}} v_4 W^\pm W^\pm \varphi^{\mp \mp} + 
\frac{g^2 v_4}{c_W} \left[ s_W^2 Z_\mu W^{+ \mu} \varphi^-_2 + \frac{\sqrt{3}}{2}(2 - s_W^2 )    Z_\mu W^{+ \mu } \varphi^-_1 + c.c. \right].
\label{eq:gauge_phi}
\end{equation}

%=====================================
\begin{figure}[!t]
  \centering
  %\begin{minipage}{1.0\textwidth}
\begin{tikzpicture}[line width=1.4 pt, scale=1,every node/.style={scale=1.0}]
\draw[fermion,black,thin] (-4.5,1) --(-3.5,0);
\draw[fermion,black,thin] (-3.5,0) --(-4.5,-1);
\draw[vector,red,thin] (-3.5,0) --(-2,0);
\draw[fermion,black,thin] (-1,-1) --(-2,0);
\draw[fermion,black,thin] (-2,0) --(-1,1);

\node at (-4.75,1.25) {$q$};
\node at (-4.75,-1.25) {$\bar{q}$};
\node at (-2.75,0.5) {$\gamma/Z$};
\node at (-0.75,-1.25) {$\Sigma^{++}$};
\node at (-0.75,1.25) {$\Sigma^{--}$};

\draw[vector,red,thin] (1,1) --(2.75,1);
\draw[fermion,black,thick] (2.75,1) --(2.75,-1);
\draw[vector,red,thin] (2.75,-1) --(1,-1);
\draw[fermion,black,thin] (4.5,-1) --(2.75,-1);
\draw[fermion,black,thin] (2.75,1) --(4.5,1);

\node at (0.75,1) {$q$};
\node at (0.75,-1) {$\bar{q}$};
\node at (3.25,0) {$\Sigma^{++}$};
\node at (5,-1) {$\Sigma^{--}$};
\node at (5,1) {$\Sigma^{++}$};

\end{tikzpicture}
%\end{minipage}
  \caption{Feynman diagrams for the production of doubly-charged fermions ($\Sigma^{\pm\pm}$) at $pp$-collider.}
\label{fig:feyn_diag_pp}
\end{figure}
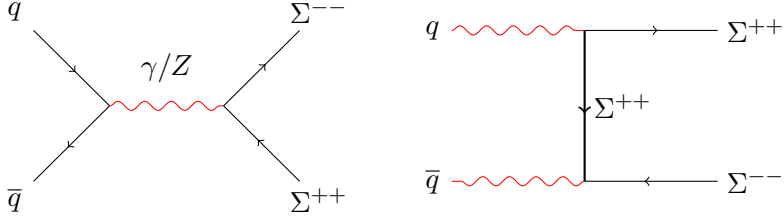
%======================================

Here, we have neglected the mixings among the individual components of the multiplets and we choose the components of 
the quintuplet fermion as well as scalar multiplets to be degenerate, which we denote as 
$M_\Sigma$ and $M_\phi$ respectively\cite{Nomura:2017abu}. We also consider 
$M_\Sigma>M_\phi$, which is naturally implied form the muon anomalous magnetic moment 
measurement, as shown in \cite{Kumar:2019tat}. For the complete Lagrangian 
we refer to  Ref:\cite{Nomura:2017abu}.
 
Further, considering the interactions in Eq. \ref{eq:yukawa}, 
$\Sigma^{\pm}$ and $\Sigma^{\pm\pm}$ can decay via the following modes,
\bea
\Sigma^{\pm} &\rightarrow &\phi_2^{\pm} \nu (\bar{\nu}) \n \\
\Sigma^{\pm} &\rightarrow &\phi^{\pm \pm} l^{\mp} \n \\
\Sigma^{\pm} &\rightarrow &\phi^{0} l^{\pm} 
\label{eqn:decayss1}
\eea
\bea
\Sigma^{\pm \pm} &{\rightarrow}&\phi^{\pm \pm} \nu (\bar{\nu}) \n \\
\Sigma^{\pm \pm} &{\rightarrow}&\phi^{\pm} l^{\pm}
\label{eqn:decayss2}
\eea
The branching ratios are assumed to be same in each decay mode of $\Sigma^{\pm}$ and $\Sigma^{\pm\pm}$. 
Interactions in Eq. \ref{eq:gauge_phi} allow the decays of the charged scalars into 
SM gauge bosons $viz$,
\bea
\phi_2^{\pm}(\phi_1^{\pm})&\rightarrow & W^{\pm} Z \n \\
\phi^{\pm \pm}~~~~ &\rightarrow & W^{\pm} W^{\pm} \n \\
\phi^{0}~~~~~~ &\rightarrow & W^{+} W^{-}.
\label{eqn:phidecay}
\eea
In the following sections, we  perform a collider study of the $\Sigma^{\pm}$ and $\Sigma^{\pm\pm}$ when they 
decay via the charged scalars \footnote {We denote $\phi_2^{\pm}= \phi^{\pm}$ in the following sections.}. 
Even though there are many possible production modes, \cite{Yu:2015pwa}, we study the pair 
production of $\Sigma^{\pm}$ and $\Sigma^{\pm\pm}$. This is because, it will be easier to reconstruct the masses
of the quintuplets if they are produced in pairs. As can be seen from the decay modes, 
the final states will have a rich collection of leptons and jets. The phenomenology of these alternative 
signatures is what we study next.  

\vspace{-0.1cm}

%====================================================================================================
\section{Phenomenology at the $p p$ Collider}
\label{sec:pp}
%=====================================================================================================
\vspace{-0.1cm}

\begin{figure}[h!]
\begin{center}
\includegraphics[width=7cm,height=6cm]{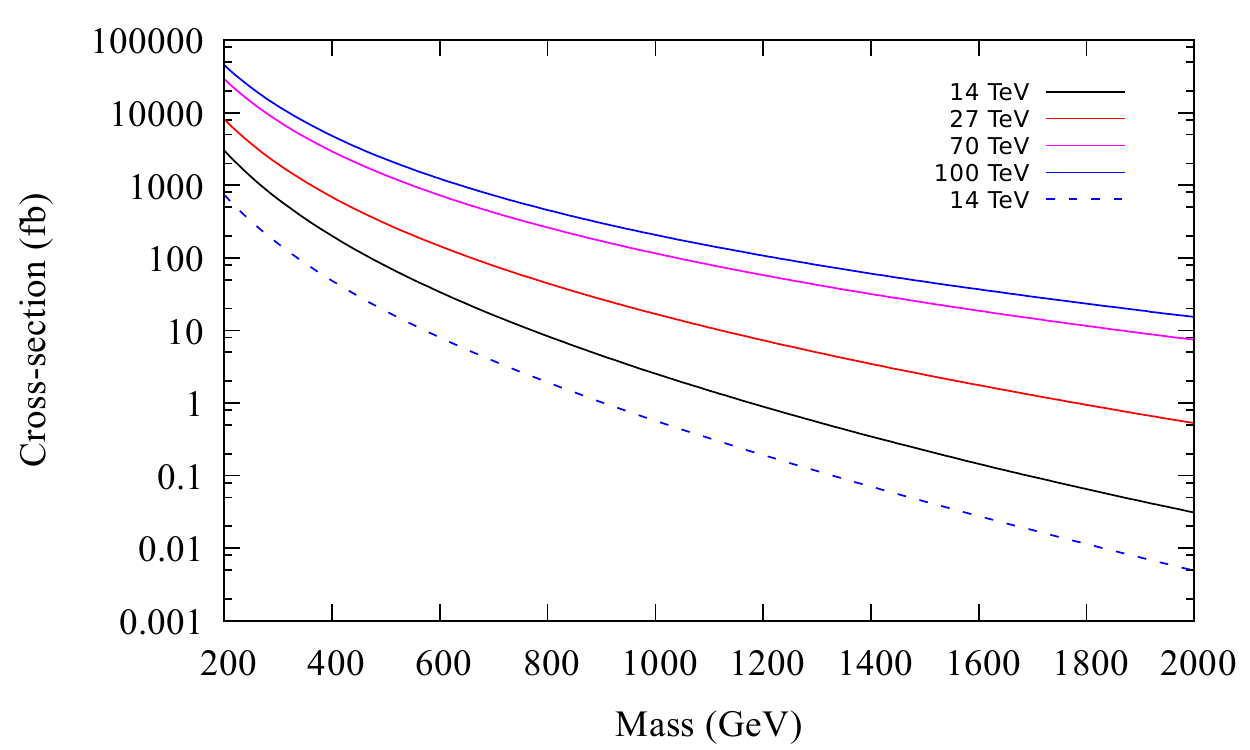}
\caption{Signal cross-section for the process, $p p \rightarrow \Sigma^{++} \Sigma^{--}$, where $p = q, \bar q, \gamma$ 
as a function of $M_\Sigma$ at different  $\sqrt s$ is shown by the solid lines. The dotted line represents the process, $p p \rightarrow \Sigma^{+} \Sigma^{-}$ at 14 TeV.}
\label{fig:ppssa}
\end{center}
\end{figure}

In this section, we discuss the collider physics of the quintuplet fermions at the LHC. 
They can be produced in $pp$ collisions through $s$ and $t$ channel processes via 
$Z/\gamma$ and $\Sigma^{\pm}$ (or $\Sigma^{\pm \pm}$), respectively. 
The cross-sections for the production of the singly charged fermions, 
$p p \rightarrow \Sigma^{+} \Sigma^{-}$, 
are smaller in comparison to those for the doubly charged fermions,
$p p \rightarrow \Sigma^{++} \Sigma^{--}$,  
as shown in Ref:\cite{Nomura:2017abu}.  
Moreover, we have included the photon-photon fusion process, hence the 
matrix element squared of pair productions are enhanced by a factor of $(Q)^4$, 
where $Q$ is the charge of the fermion. We found the
signal to background ratio to be small for the singly charged fermions. 
Hence, we choose to study the pair production of only doubly charged 
fermions at the LHC and we show the Feynman diagrams in Fig:~\ref{fig:feyn_diag_pp}. 
In Fig:~\ref{fig:ppssa}, we have shown the cross-section for the pair-production, 
$p p \rightarrow \Sigma^{++} \Sigma^{--}$, 
for different values of $\sqrt s$ at LHC where $p = q, \bar q, \gamma$. For comparison, 
cross-section for pair-production of the singly charged fermion at $\sqrt s=14$ TeV 
is also shown by the dotted curve in blue.

The inclusion of the photon PDF increases the signal 
cross-section significantly. Moreover, 
inclusion of photon PDF is important for the consistency of the calculation 
as the other PDF's are determined up to NNLO in QCD. We would like to note that, in 
view of the above, NNPDF \cite{NNPDF:2014otw,Ball:2013hta}, MRST \cite{Martin:2004dh} 
and CTEQ \cite{Schmidt:2015zda} already include photon PDF in their definitions.
In order to compute the cross-sections and generate events at the LHC, 
we incorporate the model Lagrangian in FeynRules (v2.3.13) 
\cite{Alloul:2013bka,Christensen:2008py}. Using FeynRules, we generate the 
model file for MadGraph5\_aMC@NLO (v2.2.1)~\cite{Alwall:2014hca}. For the 
cross-sections, we use {the} NNPDF23LO1 parton distributions~\cite{Ball:2012cx} 
with the factorization and renormalization scales at the 
central $m_T^2$ scale after $k_T$-clustering of the event.

%=============
%=====================

\vspace{-0.7cm}

%=======================================
\subsection{Signal}
%======================================

\vspace{-0.4cm}

Once pair-produced at the LHC, the decays of the doubly charged fermions produces the following states,
\begin{equation}
\Sigma^{++} \rightarrow \phi^{++} \nu     \rightarrow (W^{+} W^{+}) \nu; ~~~
\Sigma^{--} \rightarrow \phi^{--} \bar{\nu} \rightarrow  (W^{-} W^{-}) \bar{\nu} \n
\end{equation}
\begin{equation}
\Sigma^{++} \rightarrow \phi^{++} \nu     \rightarrow (W^{+} W^{+}) \nu;~~~ \\
\Sigma^{--} \rightarrow \phi^{-} l^{-}       \rightarrow  (W^{-} Z) l^{-} \n
\end{equation}
\begin{equation}
\Sigma^{++} \rightarrow \phi^{+} l^{+}       \rightarrow  (W^{+} Z) l^{+};~~~ \\
\Sigma^{--} \rightarrow \phi^{-} l^{-}       \rightarrow  (W^{-} Z) l^{-}
\end{equation}
with conjugate processes included in each case. 
The branching ratio in each case is assumed to be the same, as discussed in the previous section.
This gives rise to final states comprising of 
a number of leptons, jets, and missing energy, 
resulting in various multilepton, multijet and mixed states. 
After carefully analyzing each of them on 
the basis of performance over SM backgrounds and 
mass reconstruction of the doubly 
charged fermions, we decide upon two channels:
\begin{itemize}
\item Channel I: $\geq 4 \ell$ channel with two pairs of Same Sign (SS) leptons $(l^+l^+),(l^-l^-)$ + MET, where both the pairs are oppositely charged,
\item Channel II: $\geq 3 \ell$ + 2 jets channel with at least one pair of SS lepton $(l^\pm l^\pm)$+ One isolated lepton  $(l^\mp)$ + MET,
%\item (C) $\geq 2 \ell$ + 4 jets channel with atleast (1) one pair of SS lepton $(l^\pm l^\pm)$ or (2) one pair of Opposite Sign (OS) $(l^\pm %l^\mp)$.
\end{itemize}
Here, $\ell = e, \mu, \tau$. 
We also check the efficiency for the the channel $4\ell + 1/2$ jets, 
but the efficiency turns out to be less compared to channels I and II. 
For the $4\ell + 3/4$ jets channel, the signal and background cross-sections, 
both are significantly low. 
Even though it is possible to obtain a better significance 
in this channel, the number of signal events to be observed at $3000 fb^{-1}$ 
integrated luminosity is less than 10. Hence, we do not pursue this channel. 
Also, in purely leptonic channel, the leptons in the final 
states are either coming from the decay of $\Sigma^{\pm\pm}$ or 
from $W/Z$. Hence, the transverse mass of the leptons 
can be reconstructed, 
but clear mass reconstruction of the quintuplet mass is difficult 
as there are different sources of MET. However, $\geq 4\ell$ channel with 
the additional criteria as in 
channel I, predicts a good $S/B$ ratio, and hence we study it. 
The requirement of at least 4 leptons 
as well as the two SS lepton pairs in channel I makes it much cleaner 
compared to other channels, even though we do not put any restriction 
on the number of jets. We do not impose any jet or b-jet veto 
in channel-I as this will result in lesser number of signal events. Also, 
we do not go beyond the 4-lepton requirement because the signal cross-section$\times$BR falls 
off due to smaller branching ratio of $W$ and $Z$ into leptons. 
In channel II, we include both leptons and jets in the final states 
and we show that a clear mass reconstruction of $\phi^{\pm}$ or $\phi^{\pm\pm}$ 
\footnote{As they have the same mass.} and $\Sigma^{\pm\pm}$ is possible.

We choose five benchmark points (BP) in our study $viz$ 
BP1: ($M_\phi=200$ TeV, $M_\Sigma=300$ TeV)
BP2: ($M_\phi=500$ TeV, $M_\Sigma=600$ TeV), BP3:($M_\phi=700$ TeV, $M_\Sigma=800$ TeV), 
BP3:($M_\phi=900$ TeV,$M_\Sigma=1000$ TeV), BP4:($M_\phi=1000$ TeV, $M_\Sigma=1200$ TeV).
We do not study the signal for $M_{\Sigma}$ larger than 1.2 TeV 
due to two reasons. Firstly, the cross-section 
is small at a larger mass of $\Sigma$. Secondly, when $M_\Sigma > 1.2$ TeV, 
the decay products of $W$ and $Z$ bosons become colimated and the probability of 
observing them as a fatjet is larger and the analysis process will 
be very different. The fatjet scenario will also be ideal for 27 TeV com energy.

\vspace{-0.3cm}
%===========================================================================================================================================
%==========================================
\subsection{Backgrounds}
%==========================================

\vspace{-0.3cm}

The main backgrounds for channels I and II come 
from inclusive diboson production, $VV$+jets, where $V=W,Z$.
There will also be contributions from triboson 
($VVV$+jets), $HV$+jets, $t\bar t V$+jets. 
The contributions from the $t\bar t VV$+ jets and 
4-top backgrounds are found to be negligible.
For channel II, additionally, we get comparatively 
less contribution 
from $t\bar t$+ jets and $V+jets$ and  
for channel I, they are negligible. 
In \cite{Choudhury:2021nib} and the references therein, the cross-section of 
these channels has been discussed in detail.

%Unlike in other multilepton searches we can not impose a $Z/W$ veto as our signal contains them.

\vspace{-0.3cm}

%==========================================
\subsection{Collider Analysis}
%==========================================

\vspace{-0.3cm}

As a potential signature, we prefer the same sign (SS) lepton pairs 
over the opposite sign (OS) leptons, due to the abundance of the former in the signal.  
This is because of the decay of the quintuplets via the charged scalars, as shown earlier. 
On the other hand, the SM backgrounds involving one or more than one 
$Z$, are more likely to involve OS pair of leptons. Hence the signatures 
involving the SS pair of leptons suffer from less SM background. 
The signal and background are optimised over a set of selections, which we list 
in Table \ref{tab:1}. 
%------------------
\begin{table}[tpb]
\begin{small}
  \begin{center}
\begin{tabular}{|c|c|c|}
\hline
   Selections & Channel I & Channel II \\
\hline
\bf S0 & $N(\ell) \geq 4$                       &  $N(\ell) \geq 3 + N(j) \geq 2$                    \\
\hline
\bf S1 & $p_T(l)> 30$ GeV                            & $p_T(l)> 30$ GeV            \\
                 & $|\eta|(l) < 2.5$                &  $|\eta|(l) < 2.5$            \\
                 & $\Delta R_{\ell \ell/j} > 0.3$    & $\Delta R_{\ell \ell/j}> 0.3$            \\
                 &                                   & $p_T(j)> 30$ GeV            \\
                 &                                  &  $|\eta|(j) < 2.5$          \\
                 &                            &  $\Delta R_{jj} > 0.3$       \\
\hline        
\bf S2     &  $S_T(l) > 400 \gev$ & $S_T(l) > 200$ GeV \\
            & $M (\ell^+, \ell^+), M(\ell^-, \ell^-) > 100\gev$ &  $M (\ell^\pm, \ell^\pm) > 100\gev$        \\   
                              
            &$\Delta R(\ell_0,\ell_1) > 1.5$   &$\Delta R(\ell,\ell) > 1.5$\\
             &                &    $H_T(j) > 300$ GeV                       \\
\hline
 \bf S3     &  &  $60 \gev <M(j, j) < 120\gev$ \\
                  &                  &   $(M_\phi-100)<M(j,j,\ell)<(M_\phi+100)$ \\
\hline

\end{tabular}
\caption{\label{tab:1} Selections $S1$, $S2$ and $S3$ for channels I and II.}
\end{center}
\end{small}
\end{table}
%------------------------------------------------

\begin{figure}
\begin{center}
\includegraphics[width=5.5cm,height=4.5cm]{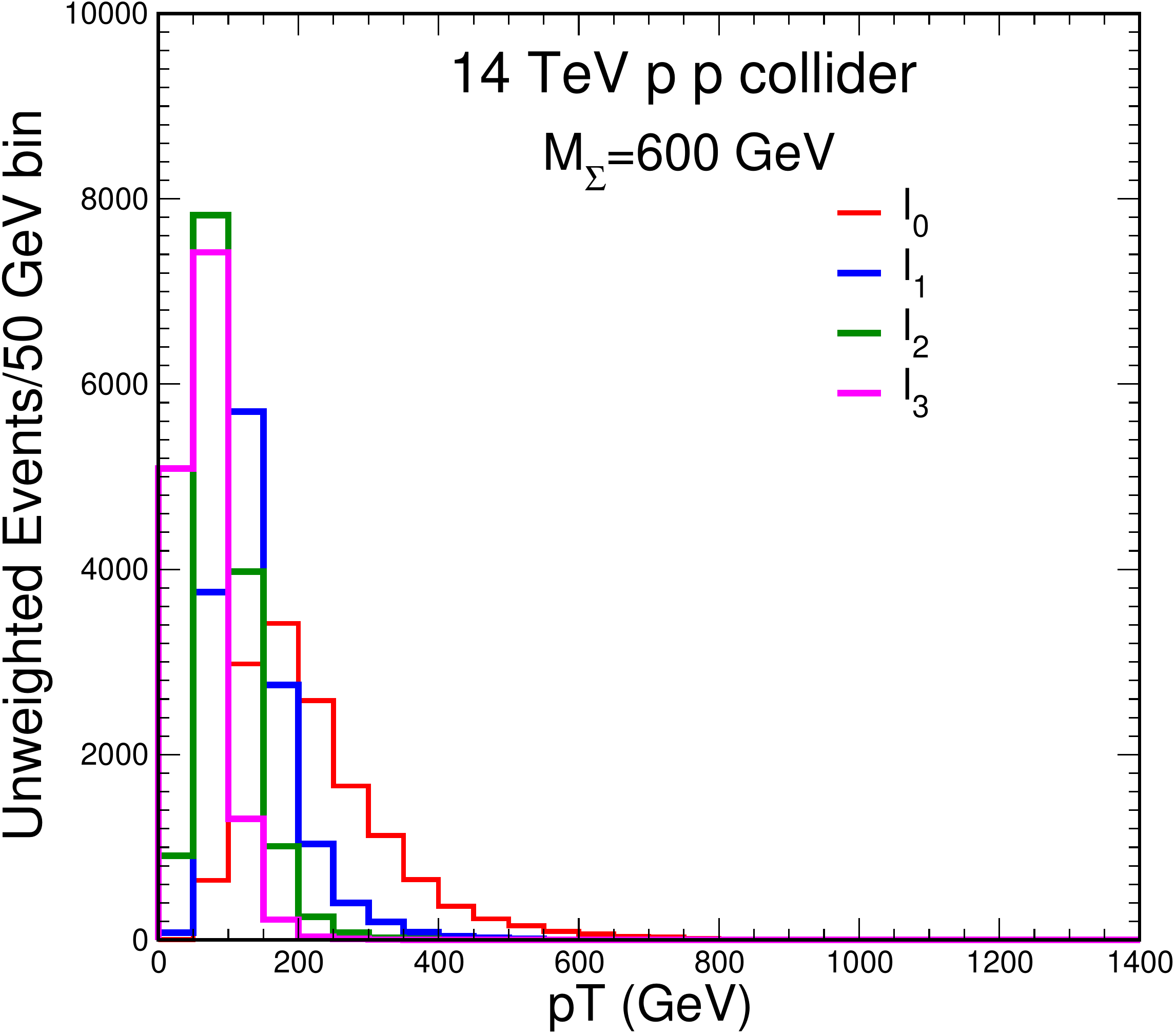}
\includegraphics[width=5.5cm,height=4.5cm]{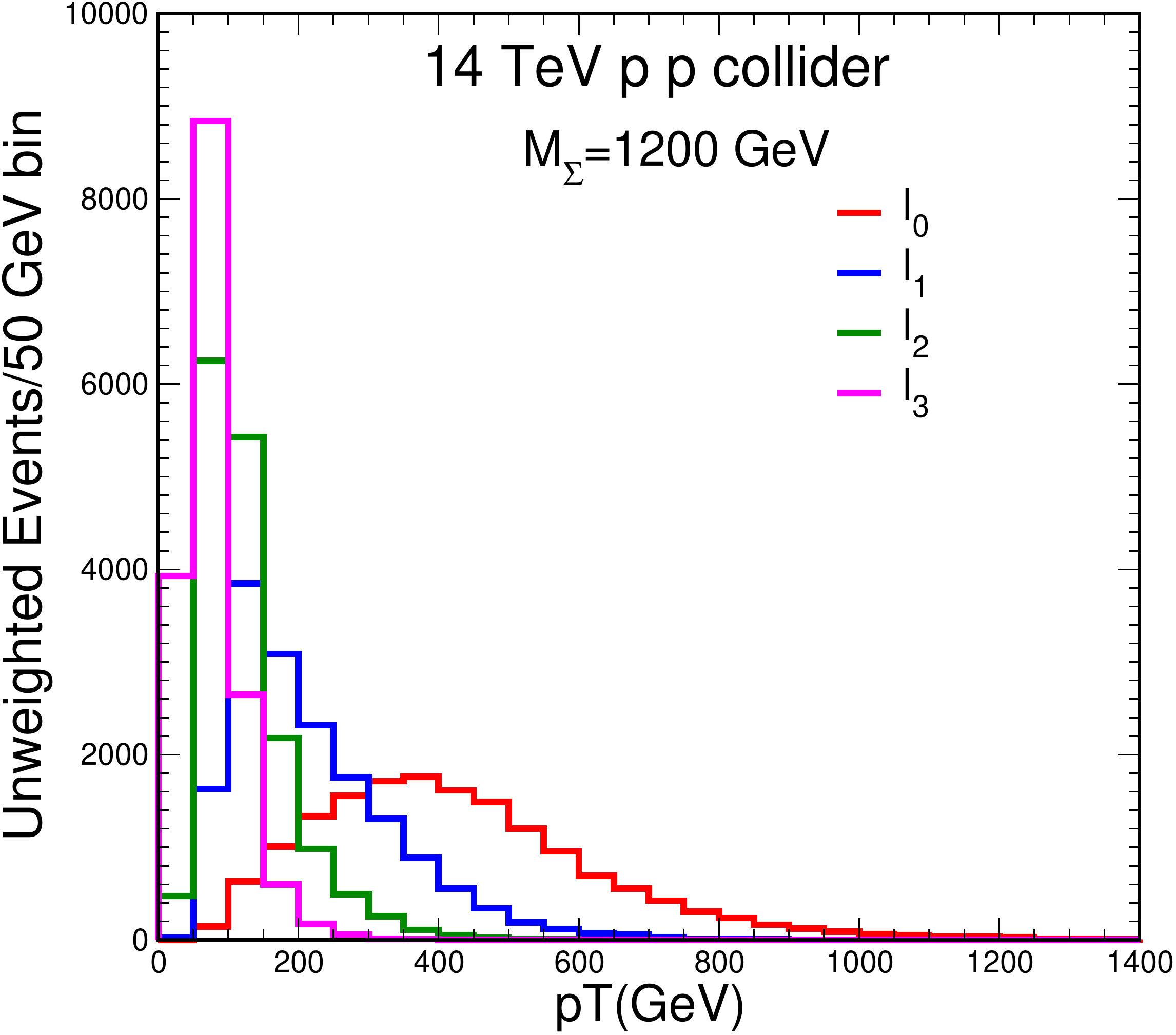}\\
\includegraphics[width=5.5cm,height=4.5cm]{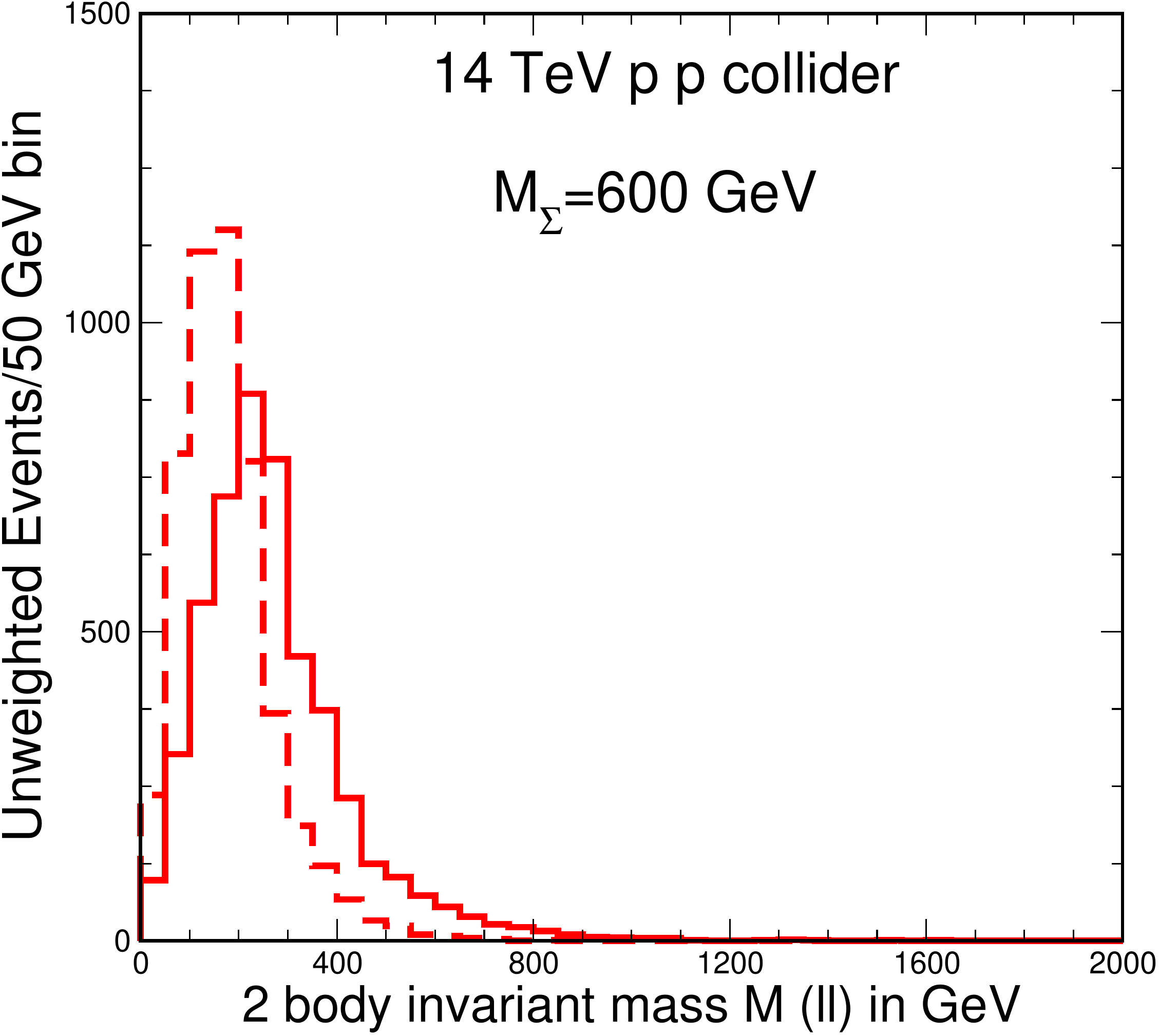}
\includegraphics[width=5.5cm,height=4.5cm]{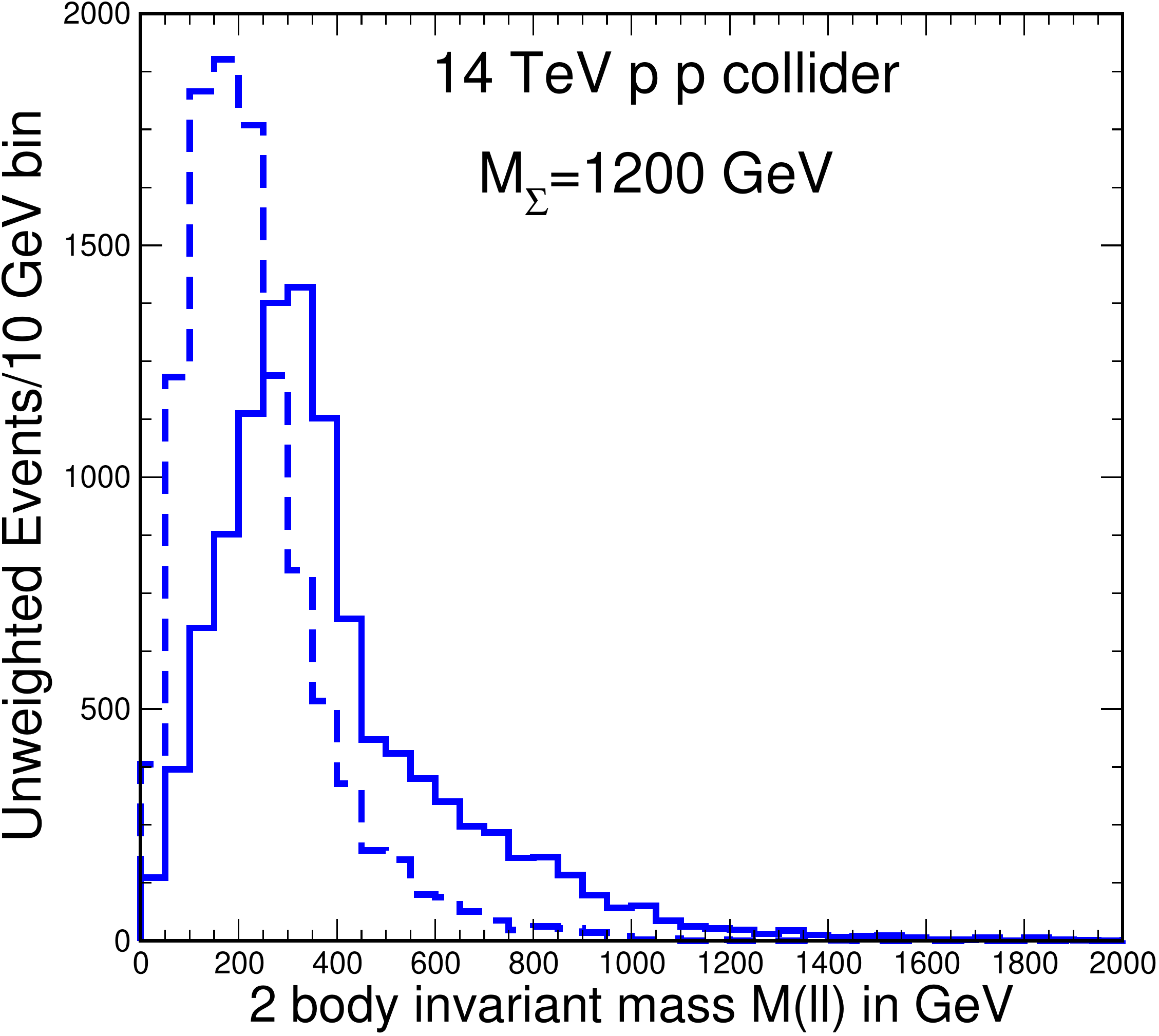}
\caption{(Top) Distributions of $pT(l)$ and (Bottom) same sign lepton pair invariant mass distributions for $M_{\Sigma}=$ 600 GeV and 1200 GeV (BP2 and BP5) respectively in channel I at 14 TeV LHC. The solid and dashed lines correspond to the first and second pair of SS lepton pairs.}
\label{fig:dist1}
\end{center}
\end{figure}

\vspace{-0.2cm}
%--------------------------------------------------------------------
\paragraph{(I) \underline {$\geq 4\ell$ channel with $(l^+l^+)$ and $(l^-l^-)$ pair+MET:}}
%--------------------------------------------------------------------
In the multilepton searches performed for exotic particles 
(vector like leptons, charged scalars etc.), 
only one OS or SS pairs are identified in most of the cases, the only exception being 
the searches for multicharged scalars \cite{ATLAS:2017xqs}. 
This is largely due to the 
standard decay modes of the BSM particles. Whereas, in channel I, we have the 
requirement for two SS lepton pair, with pairwise opposite charge. In channel I, 
the quintuplet fermions 
decay to $Z$ and/or $W$ via the charged scalars, and most of the final state leptons 
come from the decay of the $Z/W$ bosons. Hence, we do not apply any $Z/W$ veto. 
We arrange the leptons ($\ell_i$) in the descending order of $pT$. 
In this channel, the SS leptons appear directly 
from the direct decay of $\Sigma^{\pm\pm}$ and from the decay of 
$W/Z$ decay in the same decay chain.
We plot the lepton's $pT$ and the invariant mass distribution of 
the SS lepton pairs for 2 BP's (BP2 and BP5) in Fig:~\ref{fig:dist1}. 
The solid line represents the invariant mass distribution if at least one SS 
pair is present, and the dotted line represents the same for the 
additional SS pair, where these two pairs have 
opposite charge. Based on the two body mass distribution, we have imposed 
the selection $M (\ell^\pm, \ell^\pm) > 100\gev$. 
%%%%%%%%%%%%%%%
%%%%%%%%%%%%%%%%%%%%%
%\vspace{-0.5cm}

\begin{figure}
\begin{center}
\includegraphics[width=5.5cm,height=4.5cm]{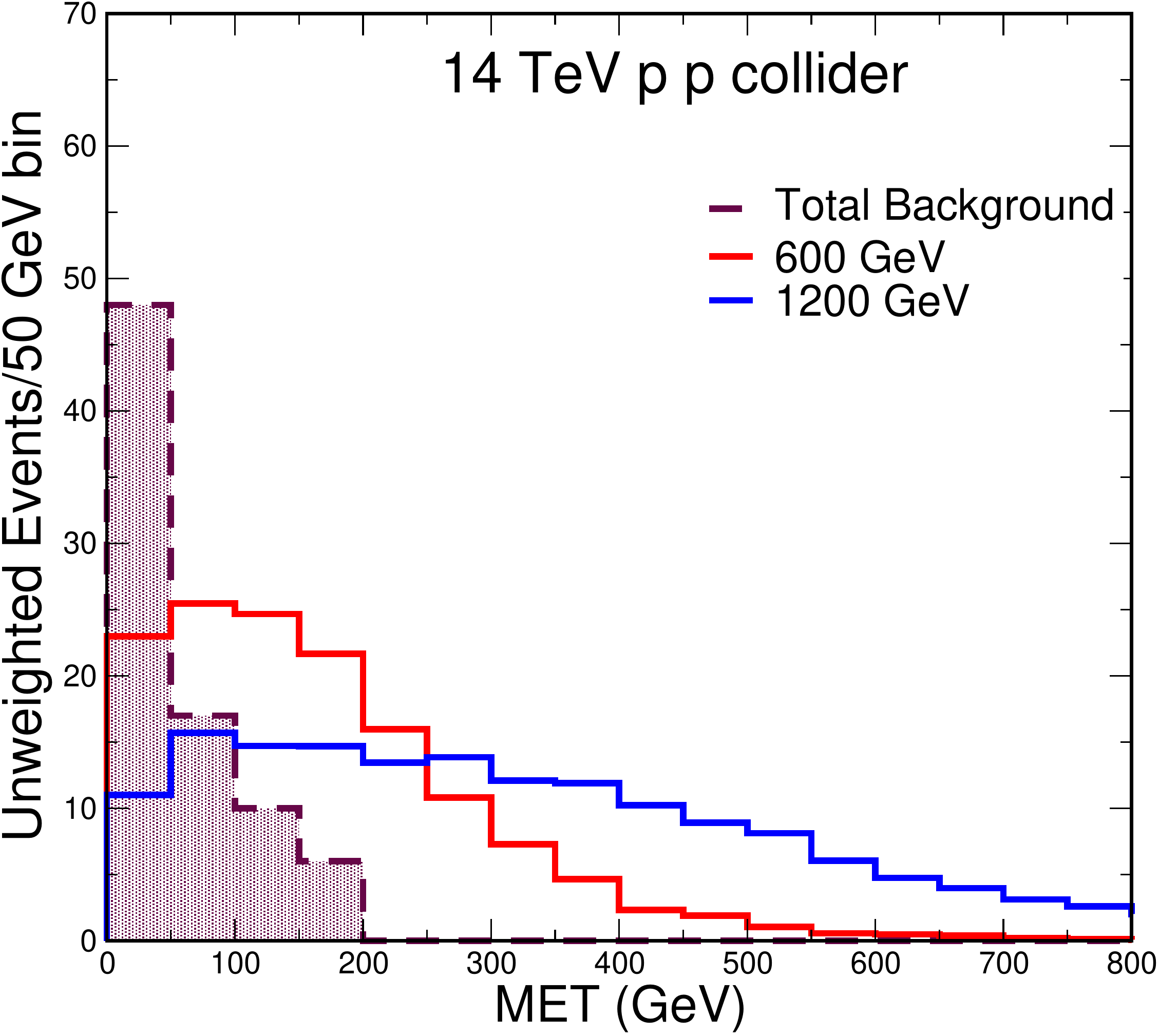}
\includegraphics[width=5.5cm,height=4.5cm]{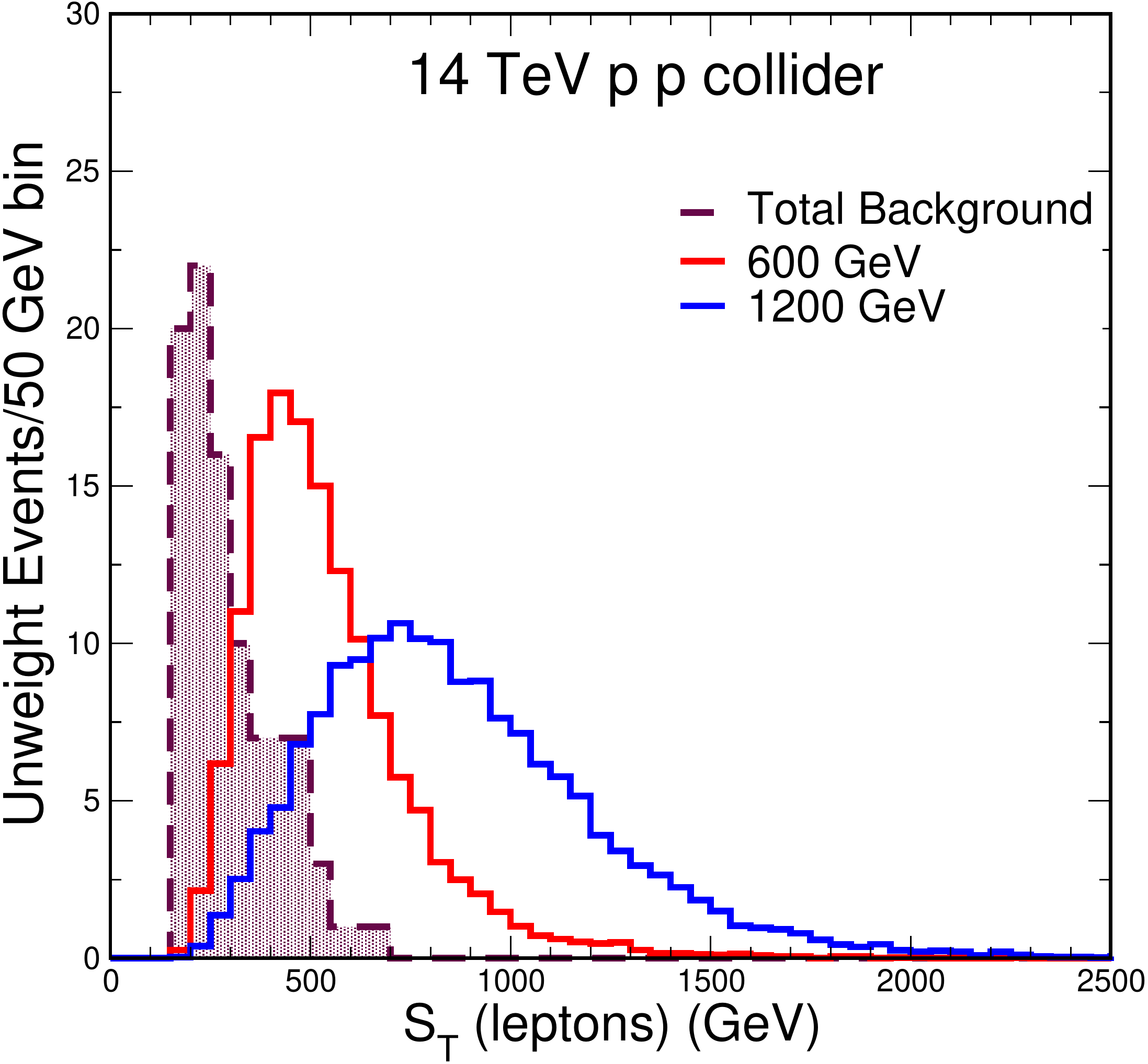}
\caption{(left) Transverse missing energy (MET) and (right) sum of lepton $pT$ ($S_T(\ell)$) 
distribution for $M_{\Sigma}=$ 600 GeV and 1200 GeV, and 
total background (shadowed region) in channel I.}
\label{fig:dist2}
\end{center}
\end{figure}

\begin{figure}
\begin{center}
\includegraphics[width=5.5cm,height=4.5cm]{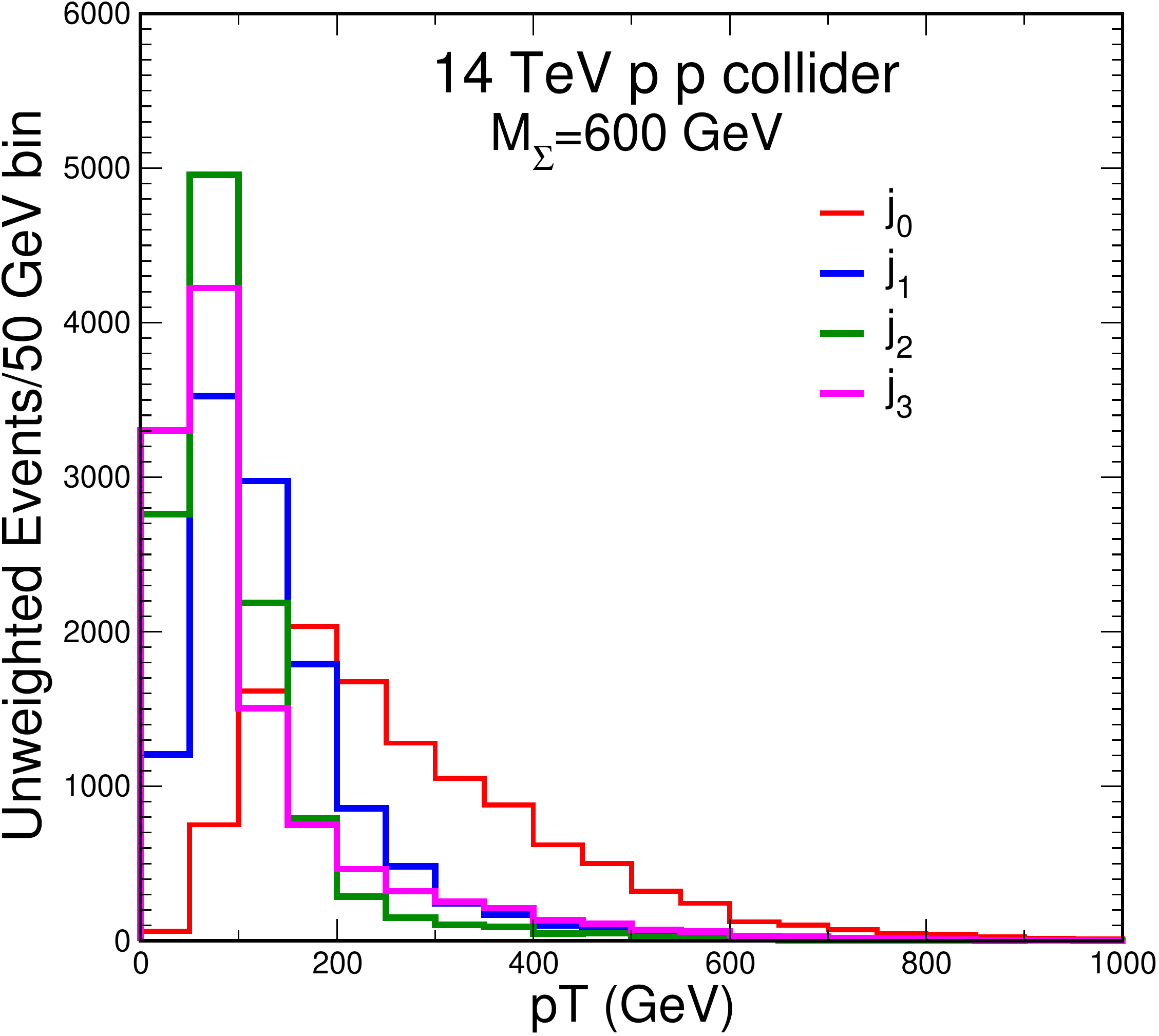}
\includegraphics[width=5.5cm,height=4.5cm]{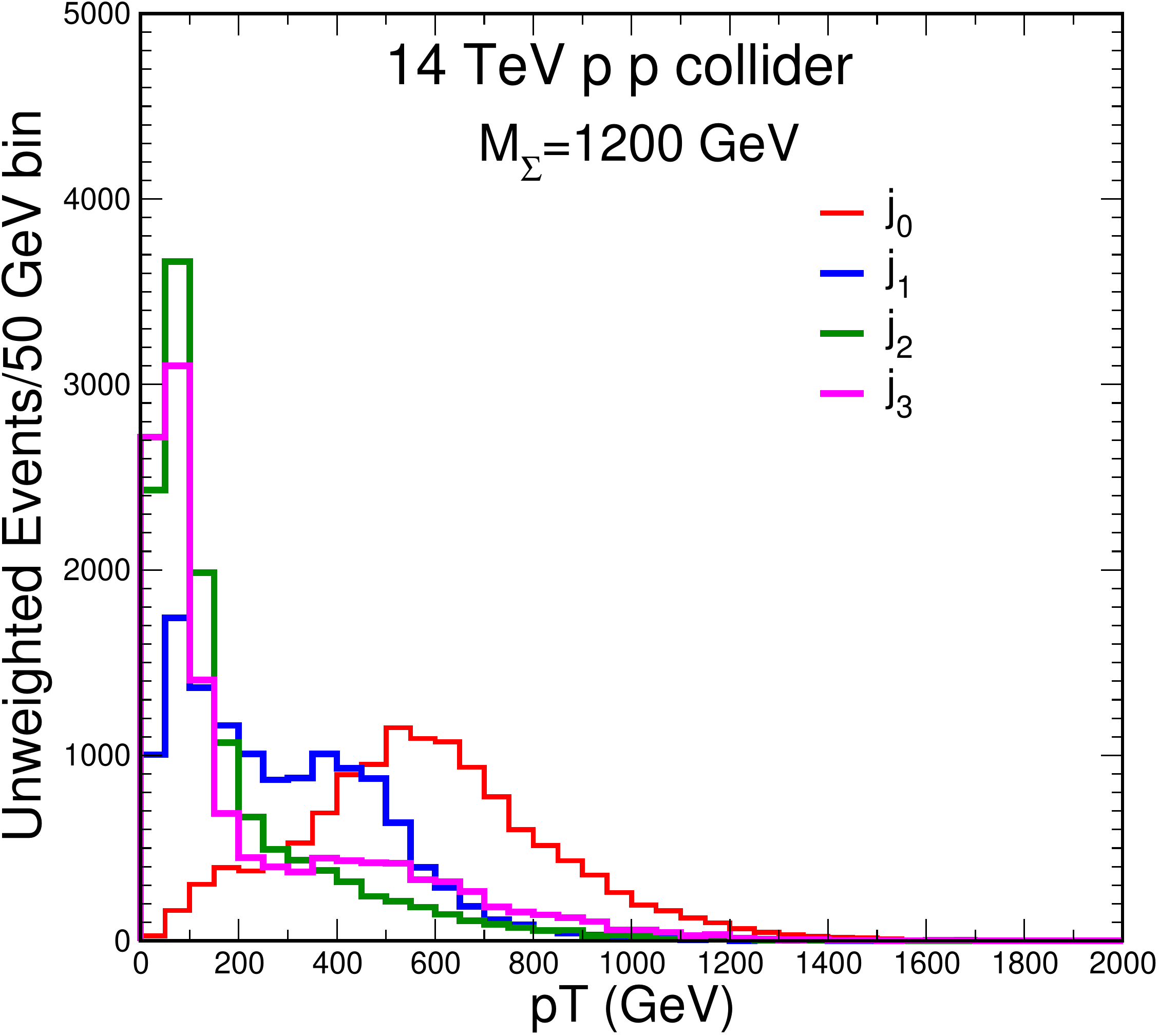}\\
\includegraphics[width=5.5cm,height=4.5cm]{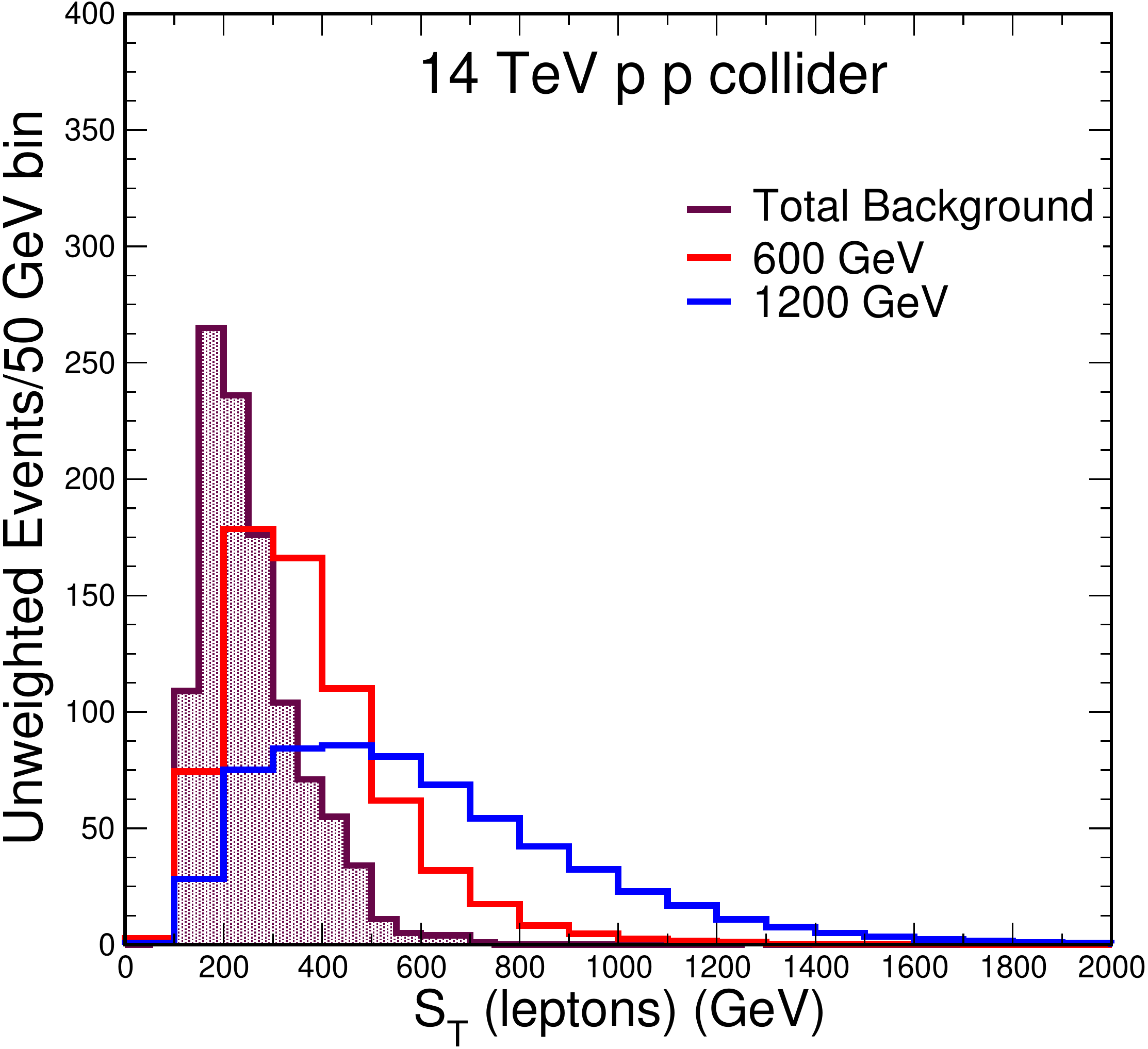}
\includegraphics[width=5.5cm,height=4.5cm]{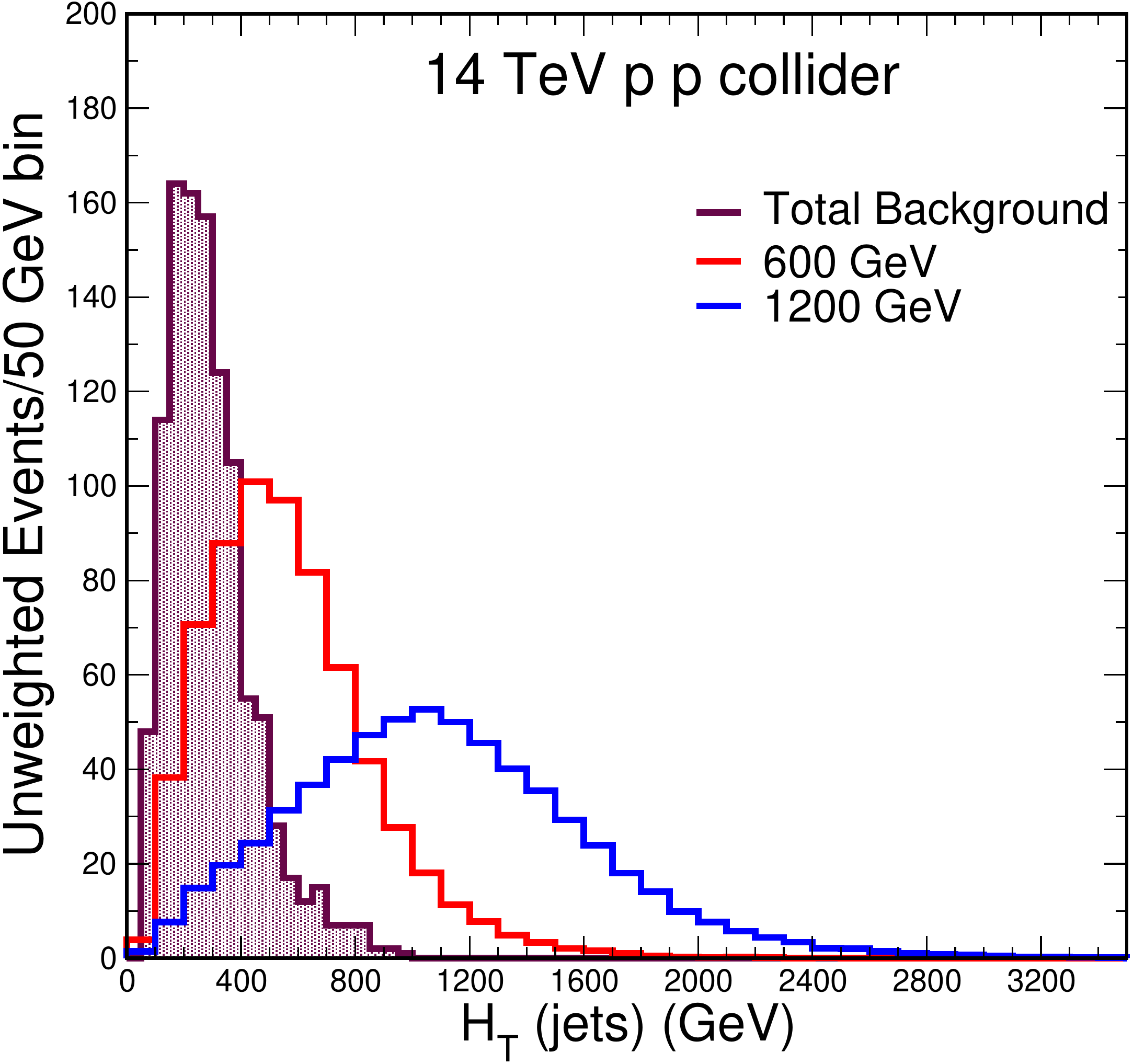}
\caption{
(Top) Distributions of jet $pT$ and (Bottom) sum of lepton $pT$ ($S_T(\ell)$) (left) and sum of jet $pT$ ($H_T(j)$) (right), for $M_{\Sigma}=$ 600 GeV and 1200 GeV, respectively, for channel II at 14 TeV LHC. The shadowed region corresponds to the total background.}
\label{fig:dist3}
\end{center}
\end{figure}

In Fig:~\ref{fig:dist2}, 
we plot the sum of lepton $pT$ ($S_T (\ell)$) and 
missing transverse energy (MET) distributions for the signal and the total background. 
Note that, a substantial amount of 
MET is present in the signal, as well as in the SM background. 
Hence, we restrain from putting any cut on MET in order to get most of the signal events.
Even though the signal has a very high $S_T (\ell)$ compared to the background, 
the set of selections optimise for $S_T (\ell)> 400$ GeV for the whole 
signal region under consideration. For example, if we focus on the region $M_\Sigma > 1$ TeV only, 
$S_T  (\ell)> 600$ GeV gives a much better $S/B$ ratio. But wwe choose to use only one value for the 
$S_T  (\ell)$ selection for our BP points.
Based on the plots of the kinametic variables, we optimize the selections at the given 
values in Table \ref{tab:1}. We found that 
the cut on $S_T(\ell)$ 
is sufficient to suppress the background in channel I. 
Moreover, the leptons with heighest $pT$ will have a large separation 
compared to the other leptons, as they are form the separate 
decay chains of the quintuplet in most of the cases.
Thus we impose a selection on these leptons by requiring 
$\Delta R (\ell_0$,$\ell_1)>1.5$. 
%%%%%%%%%%%%%%%%%%%%%%%%%%%%%%
%%%%%%%%%%%%%%%%%%%
In Table~\ref{tab:2}, we summarize the effect of the 
selections in channel I with $(l^+l^+)$ and $(l^-l^-)$ pair + MET.
As there is no jet veto imposed, the majority of the backgrounds come from 
the diboson + jets events. Initially, this background cross section is comparably very high but after we 
impose the selection $S_2$, the background reduces further.
%We have also checked the OS criteria, that gives comparatively poor result.
%----------------------------------------------------
\begin{table}[!h]
\parbox{.5\linewidth}{
  %\begin{center}
~~~~~~~~~~~~~\begin{tabular}{|c|c|c|c|}
\hline
\hline
$M_{\Sigma}$ & $S1$ (fb) & $S2$ (fb) \\
\hline
BP1, 300 GeV& 2.335		&1.093	\\
BP2, 600 GeV&0.598		&0.219	\\
BP3, 800 GeV&0.196		&0.093\\
BP4, 1000 GeV&0.063		&0.032\\
BP5, 1200 GeV&0.024		&0.013\\
\hline										
\hline
\end{tabular}
}
%\vskip 10pt
\parbox{.5\linewidth}{
\begin{tabular}{|c|c|c|c|}
\hline
\hline
Major Backgrounds &$S1$ (fb)& $S2$(fb)\\
\hline
Di-Boson+jets& 20.42 & 0.55\\
$t\bar{t}V$       &0.25&0.07 \\
Triboson &0.082 & 0.022\\
$HV$+jets &0.048& 0.021 \\
\hline
Total   &20.80&0.66 \\
\hline
\hline
\end{tabular}
}
%\end{center}
    \caption{\label{tab:2}\em {\em (Left)} The variation of the cross-section (fb) for each of the
    BPs, as the selections are imposed at 14 TeV LHC in channel I. {\em
    (Right)} The same is shown for the background.}
\end{table}
%%%%%%%%%%%%%%%%%%%%%%%%%%%%%%%%%
%---------------------------------------------------------------------------------------------
\paragraph{(II) \underline {$\geq 3\ell$ Channel with $(l^\pm l^\pm)$ pair + $l^\mp$ + $\geq 2$ jets channel}:}
%---------------------------------------------------------------------------------------------
In this channel, we require the presence of at 
least three leptons as well as two or more jets. 
In Fig:~\ref{fig:dist3}, we show the $p_T$ distributions 
of the jets, and also the sum of $pT$ 
for the jets ($H_T(j)$). The $pT$ distribution of the 
leptons are mostly the same as channel I but 
as the number of leptons in channel II  
is less than channel I, the 
$S_T(\ell)$ distribution peaks at a lower value 
compared to channel I. In channel II, 
we identify at least one SS lepton pair in a manner as stated in channel I.
The selections in channel II are summerized in Table \ref{tab:1}. 
Additionally, 
we find that a cut on the minimum value of $H_T(j)$ is useful 
to minimise the background. 

%---------------------------------
%---------------------------------

\begin{figure}
\begin{center}
\includegraphics[width=5.5cm,height=4.5cm]{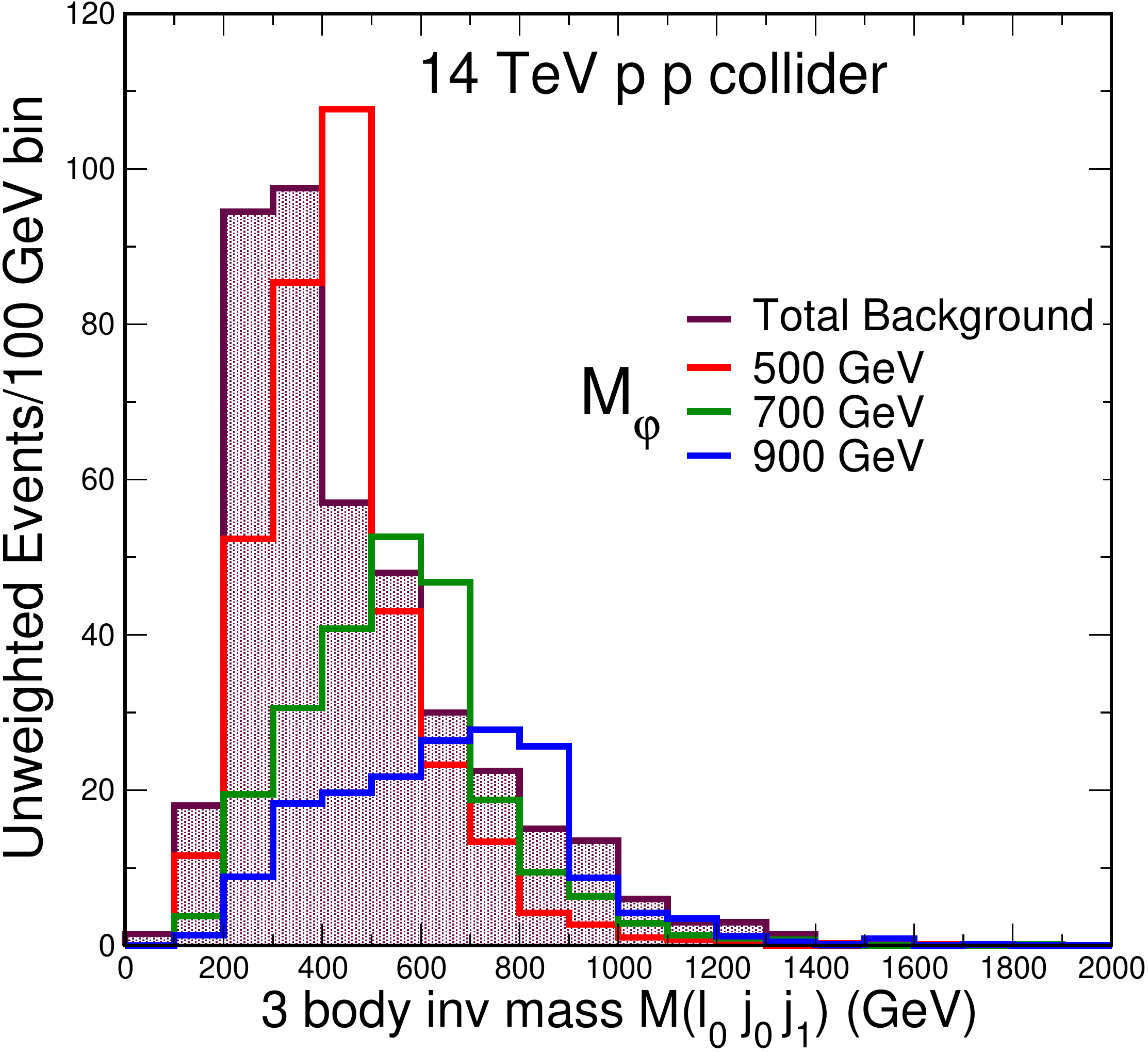}
\includegraphics[width=5.5cm,height=4.5cm]{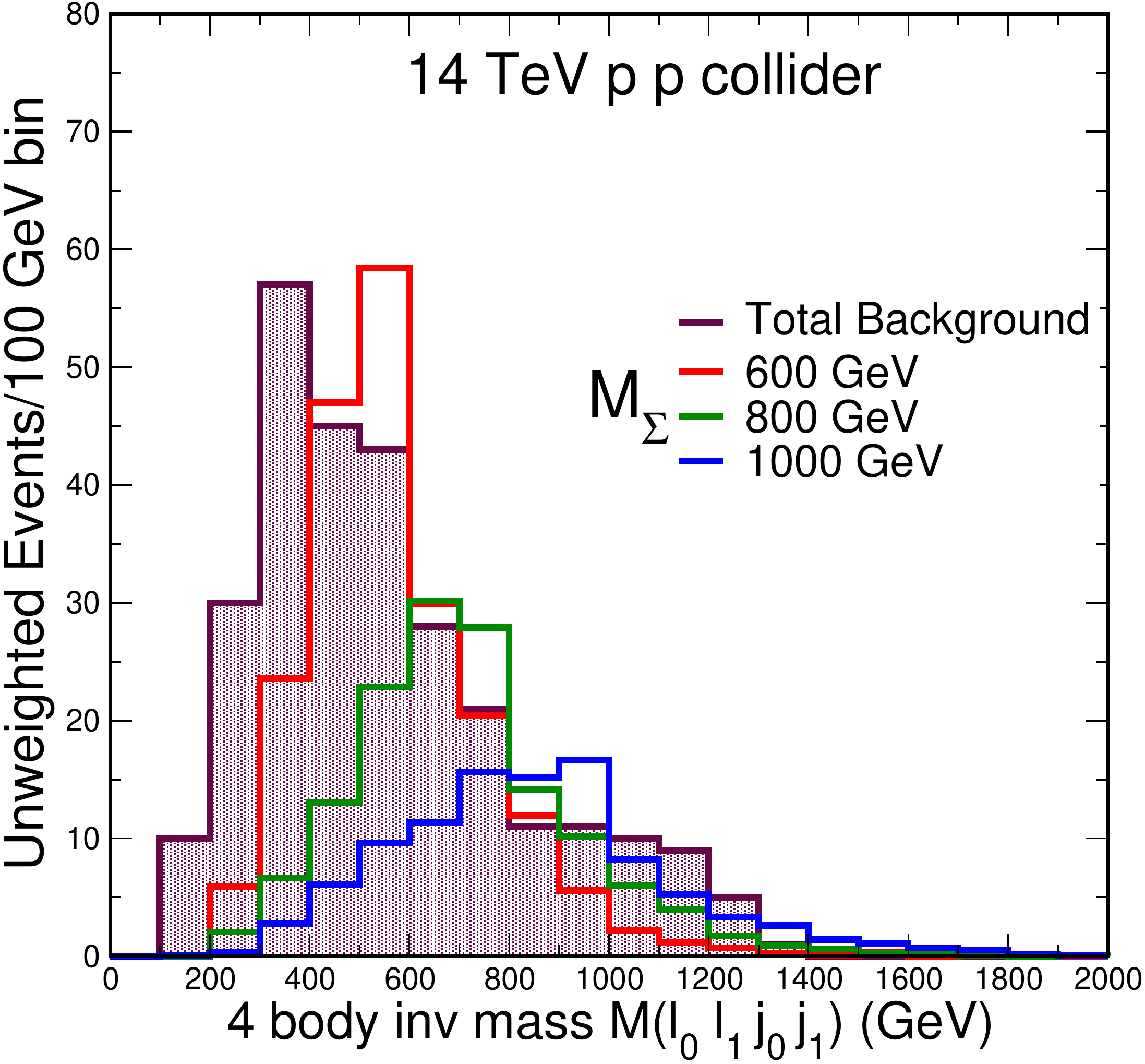}
\caption{(left) Three body 
invariant mass $M(\ell\ell jj)$ and (right) four body invariant mass $M(\ell\ell jj)$ for $M_{\phi}=$ 500, 700, 900 GeV and $M_{\Sigma}=$ 600, 800, 1000 GeV, respectively, for channel II at 14 TeV LHC. The shadowed region corresponds to the total background.{\bf change plot}}
\label{fig:dist4}
\end{center}
\end{figure}

The main objectives in channel II are to 
construct the three and four body 
invariant mass distribution, $M(\ell \ell j)$ and 
$M(\ell\ell jj)$, for the reconstruction of 
$M_\phi$ and $M_\Sigma$, respectively. Note that, this is only possible 
when we consider the decay of the quintuplet via the singly charged scalar.
Even though it is theoretically possible to reconstruct the mass of the quintuplet from 
$M(\ell jjj)$ also, it is harder to select the exact jets for the distribution. 
Hence, we consider 
the case when $W$ and $Z$ decay through leptonic mode and hadronic mode respectively.
At first, we select two SS leptons in such a way that they must come from the same decay chain. 
One lepton is coming from the decay of the quintuplet and the another is from the $W$. We demand 
$\Delta R(\ell\ell) >1.5$ for these two leptons.
Then we select two jets coming from the the decay of the $Z$ boson, 
by requiring $60 < M(j j) < 120$ GeV.
The three body mass distribution $M(\ell \ell j)$ 
and the four body mass distribution 
$M(\ell\ell jj)$ reconstruct the masses of $\phi$ and $\Sigma^{\pm}$, which is shown 
in Fig:~\ref{fig:dist4}. 
%---------------------------------------------------- 
%----------------------------------------------------
We select the final events with $S_3$, where 
the events are required to satisfy the three body 
invariant mass in the window of $M_\phi\pm 100$ GeV. 
The signal and background cross-section after the cuts 
are shown in Tab~\ref{tab:3}. 
Clearly, the selection after $S_3$ gives a better signal 
to background ratio.
%---------------------------------
\begin{table}[!h]
  \begin{center}
\begin{tabular}{|c|c|c|c|}
\hline
\hline
$M_{\Sigma}$&$S1$ & $S2$ & $S3$ \\
\hline
BP1, 300 GeV& 11.55 & 4.4 & 0.112\\
BP2, 600 GeV& 2.85 &1.025 &0.028	\\
BP3, 800 GeV& 0.84 &0.39	 &0.007\\
BP4, 1000 GeV& 0.25 &0.14	 &0.0017	\\
BP5, 1200 GeV& 0.09 &0.054 &0.0005	\\
\hline										
\hline
\end{tabular}
\vskip 10pt
\begin{tabular}{|c|c|c|ccccc|}
\hline
\hline
Major Backgrounds &$S1$ & $S2$ & $S3$ (BP1)& $S3$ (BP2) & $S3$ (BP3) &$S3$ (BP4)&$S3$ (BP5)\\
\hline
Di-Boson+jets   & 45.05 &14.00 & 0.085  &5.5$\times 10^{-3}$  & 3$\times 10^{-4}$ &$6\times 10^{-5}$& $1\times 10^{-5}$ \\
$t\bar{t}V$& 10.42 &0.53  & 0.056 &1$\times 10^{-3}$    & 2$\times 10^{-4}$ &$2\times 10^{-5}$& $<10^{-5}$\\
Triboson   & 0.336 &0.013 & 0.004 & $<10^{-3}$          & $<10^{-4}$        &$<10^{-5}$       &$<10^{-6}$\\
$HV$+jets  & 1.2   &0.012 & 0.003 &$<10^{-3}$           & $<10^{-4}$        &$<10^{-5}$       &$<10^{-6}$\\
\hline  
Total   & 61.06 & 14.57 &0.148& 0.0065 &0.0005 & $7\times 10^{-5}$ & $1\times 10^{-5}$\\
\hline
\hline
\end{tabular}
\end{center}
    \caption{\label{tab:3}\em {\em (Top)} The variation of the signal cross-section (fb) for each of the
    BPs, as the selections are imposed at 14 TeV LHC. {\em
    (Bottom)} The same for the various backgrounds.}
\end{table}
%------------------------------------------------------------
%---------------------------------------------
% OS channel is checked also, poor $S/B$. More backgrounds that involves Z.
%------------------------------------------------------
%\paragraph{Other channels:}
%--------------------------------------------------
%$\geq 2\ell$ Channel with 2SS or 2OS pair + $\geq 4$ jets channel has a lot of background due to the multijet requirement.
%So, we change our focus towards the $e^{+} e^{-}$ collider, in order to analyse the channels with multiple jets.%------------------------------------------------------
\paragraph{\underline {Result}}
%-----------------------------------------------------
The significance for the discovery can be described as (see Ref: \cite{Cowan:2010js,Li:1983fv,Cousins:2007yta}),

\vspace{-0.7cm}
 
\begin{center}

\beq
\Zdis = \left [ 2\left ((s+b) \ln \left [\frac{(s+b)(b+\sigmab^2)}{b^2+(s+b)\sigmab^2}\right ]-
         \frac {b^2}{\sigmab^2} \ln\left [1+ \frac {\sigmab^2 s}{b(b+\sigmab^2)}\right ]\right)\right]^{1/2}
         \label{eq:zdis1}
\eeq

\end{center}
Where $s$ and $b$ are number of signal and background events respectively, and $\Delta_b$ is the 
uncertainty in the measurement of the background. If $\Delta_b = 0$,
\begin{center}
$\Zdis=\sqrt{2[(s+b)\ln(1+s/b)-s]}$
\end{center}
If $b$ is large,
\begin{center}
$\Zdis = s/\sqrt{b}$
\end{center}

Thus, if $b$ is small, $s/\sqrt{b}$ overestimates the significance. We 
use $\Zdis > 5$ which corresponding to $p< 2.86 \times 10^{-7}$ for different values of  
$\Delta_b$. Similarly, the significance
for exclusion is,

\vspace{-0.5cm}

\bea
\Zexc&=\left [2 \left \{ s-b \ln \left (\frac{b+s+x}{2b} \right ) - \frac{b^2}{\Delta_b^2} \ln \left (\frac{b-s+x}{2b} \right ) \right \} -
(b + s - x) (1 + b/\Delta_b^2) \right ]^{1/2} \n \\
&x = \sqrt{(s+b)^2 - 4 s b \Delta_b^2/(b + \Delta_b^2)}
         \label{eq:zdis2}
\eea

If $\Delta_b = 0$,
\begin{center}
$\Zexc = \sqrt{2(s - b \ln(1 + s/b))}$
\end{center}
%---------------------------------------------------
For 95\% confidence level (CL) exclusion
($p = 0.05$), we use $\Zexc > 1.645$ for different 
$\Delta_b $. 

\begin{figure}[tb]
\begin{center}
\includegraphics[width=6cm,height=5cm]{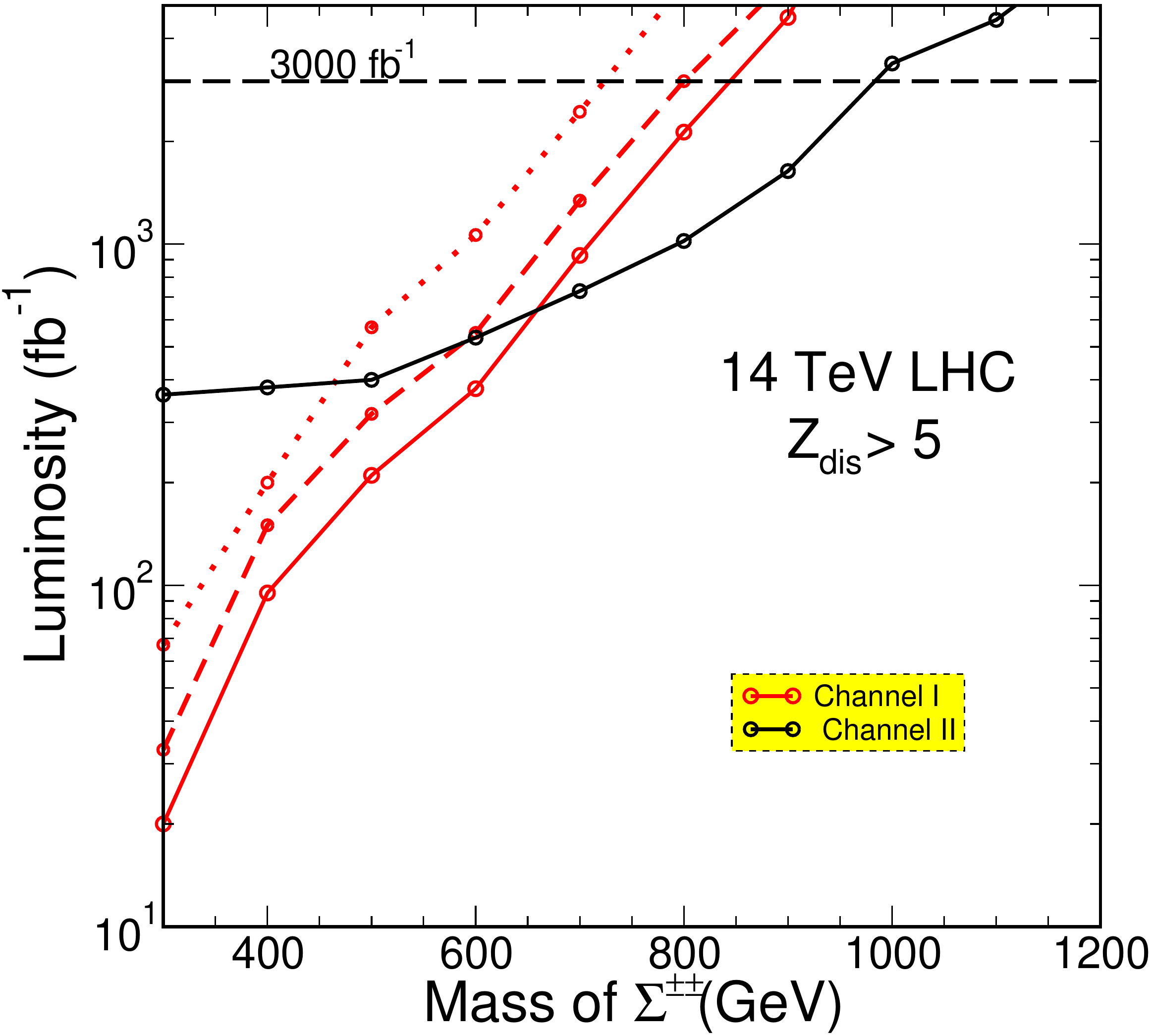}
\includegraphics[width=6cm,height=5cm]{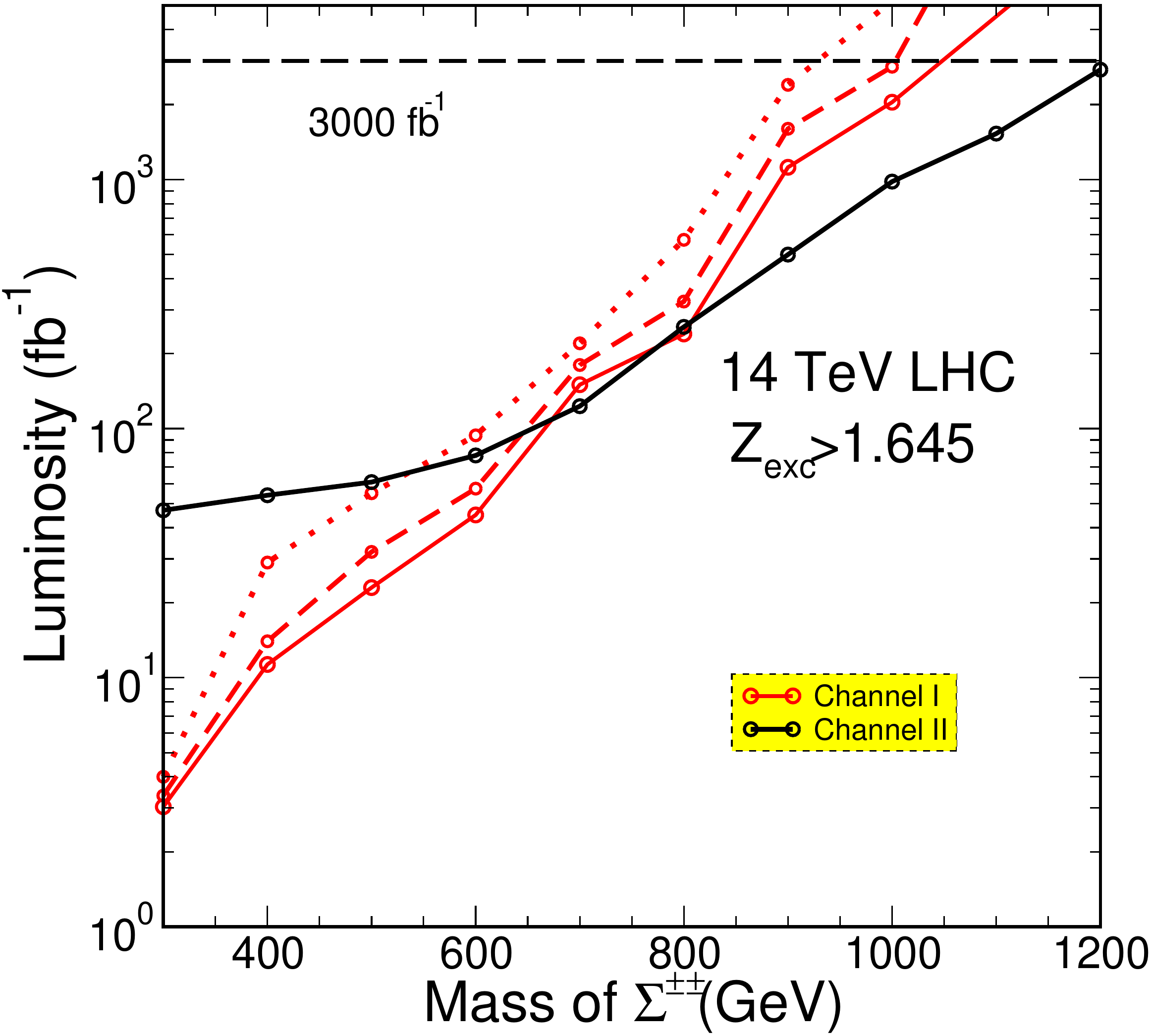}
\caption{The integrated luminosity for 
discovery (left) and exclusion (right) as a function 
of the mass of the doubly charged fermion at 14 TeV LHC. The solid, dashed and dotted lines correspond 
to 0\%, 25\% and 50\% uncertainty in the total background, respectively.}
\label{fig:result1}
\end{center}
\end{figure}

We calculate the significance using the formula in Eq.\ref{eq:zdis1},
 in order to account 
for the uncertainty in the background, as the background is small in both the channels. 
The integrated luminosity for 
discovery and exclusion as a function 
of the mass of the doubly charged fermion($M_\Sigma$) is shown in Fig:~\ref{fig:result1}.  
The prediction in 
channel I is sensitive to the uncertainty in the background, 
which we have considered to be 
$\sigma_B= 0$, $0.25\times b$, $0.5\times b$. Channel II is 
not sensitive to $\sigma_B$ as the signal 
and background cross-sections, both, are small, as given in Table~\ref{tab:3}. 
We have found that channel I and II 
have a good discovery potential for masses upto 
850 GeV and 1025 GeV respectively at 3000 fb$^{-1}$ luminosity with $\sigma_B$=0. 
In channel I, more than $3000$ fb$^{-1}$ luminosity
is required for discovery of $M_\Sigma >850$ GeV with nonzero $\sigma_B$. 

Masses upto 1.05 TeV and 1.2 TeV can 
be excluded with 95\% CL (corresponds to Z value =1.645)
at 3000 fb$^{-1}$ luminosity, with no background uncertainty. 
In channel I, with integrates luminosity of $3000$ fb$^{-1}$, 
the exclusion limit is 1 TeV and 920 GeV, 
for $\sigma_B$=0.25 and 0.5 respectively. 
Hence, we find that channel I and II have a good prospect 
for both exclusion and discovery of the doubly 
charged fermions in HL-LHC, with $3000$ fb$^{-1}$, with the added advantage of mass reconstruction 
for the doubly charged fermion and the charged scalar in channel II. 
%===================================
%--------------------------------------------
%=================================================================================
\section{Phenomenology at the $e^{+}e^{-}$ Collider}
\label{sec:ee}
%=================================================================================
\vspace{-0.1cm}

We have shown in Sec:~\ref{sec:pp} that the pair production 
cross-sections of the singly charged fermions ($\Sigma^{\pm}$) 
are smaller compared to the doubly charged fermions ($\Sigma^{\pm\pm}$)  
at p-p collision, where both are components of a fermionic quintuplet. The small cross-section makes it difficult 
to observe singly charged fermions at 14 TeV LHC when 
we look for the alternative signatures in our 
model. Even increasing the center of mass 
energy further upto 27 TeV does not solve the issue.
These singly and doubly charged fermions can also be produced 
in linear colliders, such as the $e^+e^-$ collider, which in turn generate 
multiple leptons and jets in the final state. Even though it is possible to 
observe alternative signatures for both singly and doubly charged fermions 
at  the $e^+e^-$ colliders, we restrict ourself to the case 
of the singly charged fermion. The production of the doubly charged fermions 
lead to more leptons and jets in the final state 
than the singly charged fermions. Here, we choose to study the final states once 
the singly charged fermions are produced in pair at $e^+e^-$ collider. The analysis 
for the doubly charged fermions will be similar to this.
Moreover, at LHC, being a $pp$ collider, the multijet signals are complicated to study due to the heavy 
QCD backgrounds. But the $e^+e^-$ collider offers a much clean environment. Hence, 
the SM background for the signal involving multiple jets are 
remarkably small compared to $pp$ collider. 

\vspace{-0.3cm}
%-----------------------------------------------------------
\subsection{Signal}
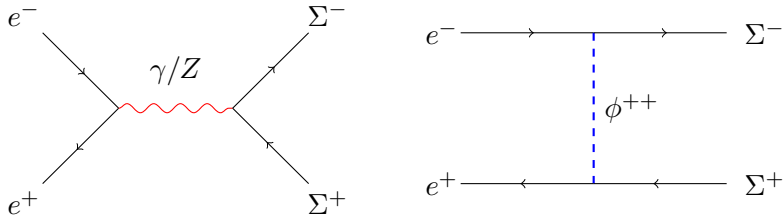
\begin{figure}[!t]
  \centering
  %\begin{minipage}{1.0\textwidth}
\begin{tikzpicture}[line width=1.4 pt, scale=1,every node/.style={scale=1.0}]

\draw[fermion,black,thin] (-4.5,1) --(-3.5,0);
\draw[fermion,black,thin] (-3.5,0) --(-4.5,-1);
\draw[vector,red,thin] (-3.5,0) --(-2,0);
\draw[fermion,black,thin] (-1,-1) --(-2,0);
\draw[fermion,black,thin] (-2,0) --(-1,1);

\node at (-4.75,1.25) {$e^-$};
\node at (-4.75,-1.25) {$e^+$};
\node at (-2.75,0.5) {$\gamma/Z$};
\node at (-0.75,-1.25) {$\Sigma^+$};
\node at (-0.75,1.25) {$\Sigma^-$};

\draw[fermion,black,thin] (1,1) --(2.75,1);
\draw[scalar,blue,thick] (2.75,1) --(2.75,-1);
\draw[fermion,black,thin] (2.75,-1) --(1,-1);
\draw[fermion,black,thin] (4.5,-1) --(2.75,-1);
\draw[fermion,black,thin] (2.75,1) --(4.5,1);

\node at (0.75,1) {$e^-$};
\node at (0.75,-1) {$e^+$};
\node at (3.25,0) {$\phi^{++}$};
\node at (5,-1) {$\Sigma^+$};
\node at (5,1) {$\Sigma^-$};

\end{tikzpicture}
%\end{minipage}
  \caption{Feynman diagrams for the production of singly charged quintuplet fermion at the $e^+e^-$ collider.}
\label{fig:feyn_diag}
\end{figure}

%--------------------------------------------------------
\vspace{-0.3cm}
The singly charged fermions($\Sigma^{\pm}$), 
can be produced in pairs at the $e^+e^-$-collider via the 
gauge couplings as described in the previous section. 
The Feynman diagrams for the pair production are shown in 
Fig:~\ref{fig:feyn_diag}. In general, the process proceeds through $s$-channel
via $\gamma$ and $Z$ boson exchange. But, in this particular model, 
there is an extra contribution 
coming from the $t$-channel diagram via the doubly charged scalar. 
The cross-section due to the t-channel diagram is large compared to the 
other diagram. 
However, the contribution in the total cross-section is not so large 
due to destructive interference between the $s$- ans the $t$-channel diagrams.
%-----------------------------
%-----------------------------
The effect of the polarization of the electron and 
positron beam has been discussed in detail 
in Ref: \cite{Shang:2021mgn}, 
and we have followed the exact same polarization of the $e^+$ and $e^-$ beam
which leads to maximum -60\% left right asymmetry ($A_{LR}$).
 
\begin{figure}[h!]
\begin{center}
\includegraphics[width=5.3cm,height=5cm]{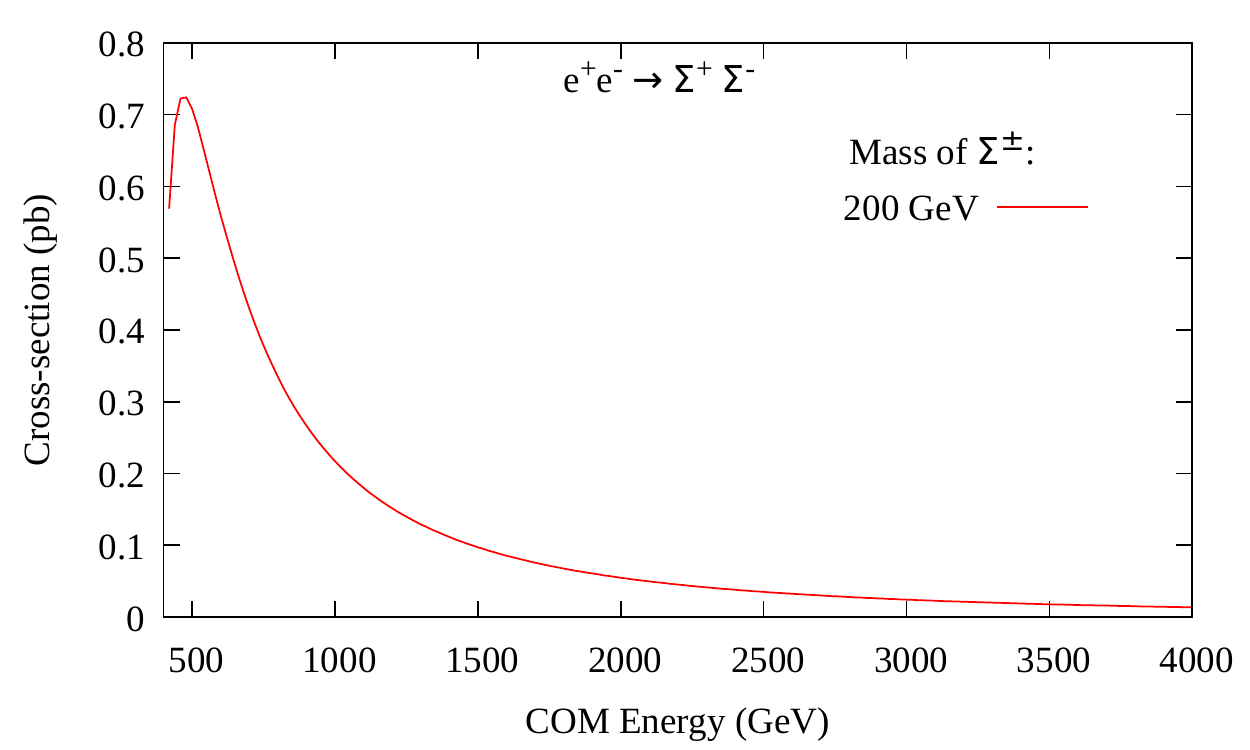}
\includegraphics[width=5.3cm,height=5cm]{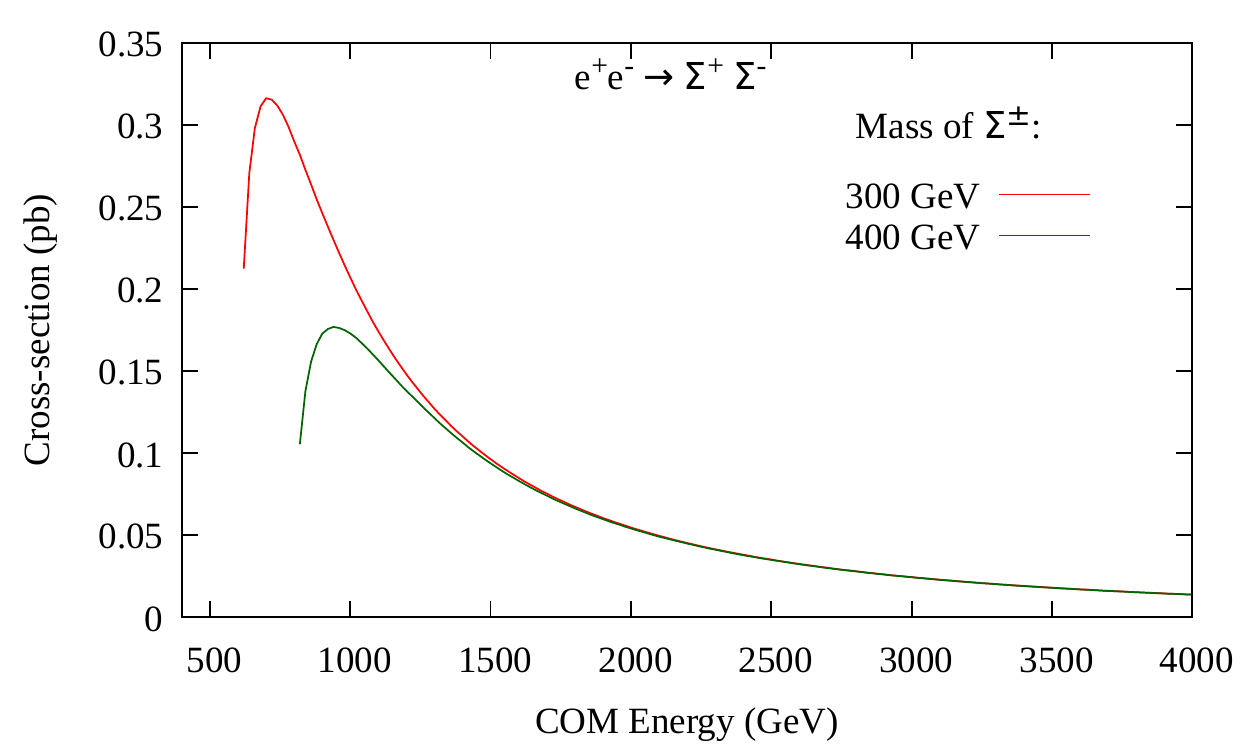}
\includegraphics[width=5.3cm,height=5cm]{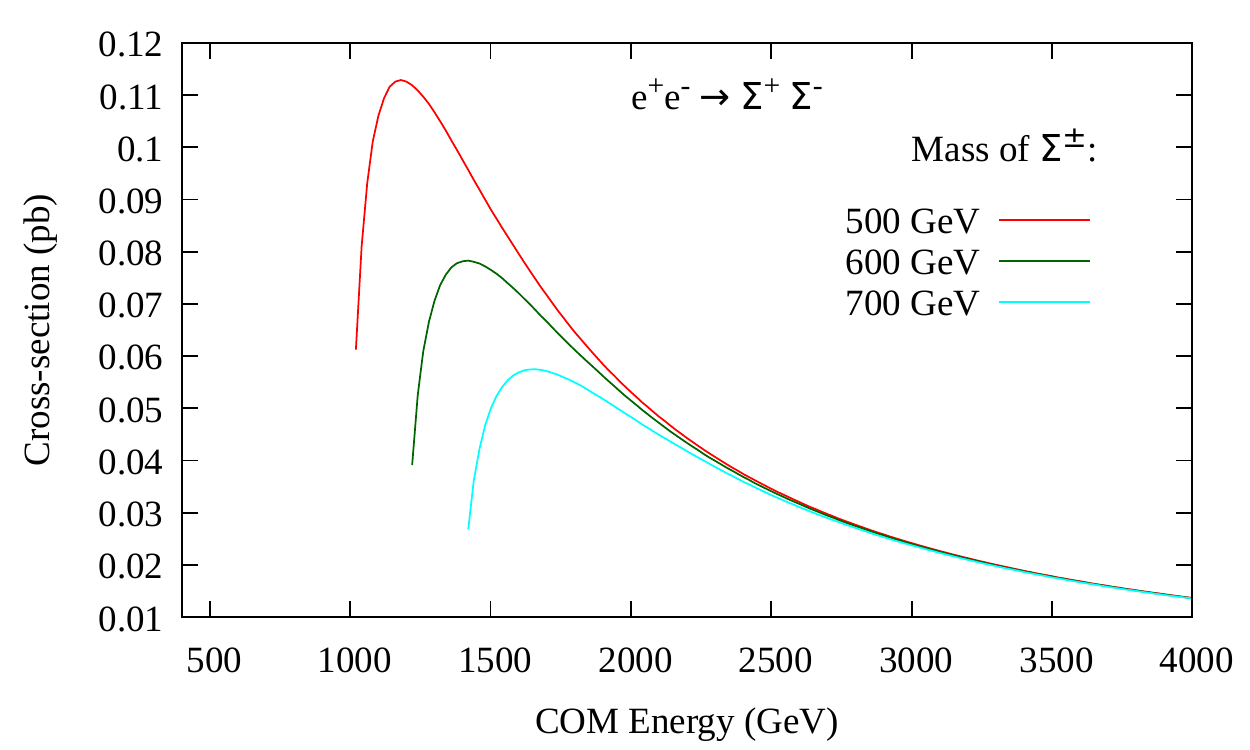}
\caption{Pair production cross-sections for the singly charged fermions of different masses  
as a function of center of mass energy, at $e^+e^-$ collider.}
\label{fig:prod1}
\end{center}
\end{figure}

%in (ref). We choose the polarization as, 
%($Pe^−= 0.8, Pe^+ = −0.3$) for $\sqrt s$ = 500 GeV, $Pe^−= 0.8, Pe^+ = −0.2$ for 
%$\sqrt s$ = 1000 GeV, and  $Pe^−= 0.8, Pe^+ = 0$ for 
%$\sqrt s = 1500$ GeV, which leads to maximum
%-60\% left sight asymmetry ($A_{LR}$), which is due to 
%the chirality-violating effect caused by the $Z$
%boson. 
The production cross-sections are computed in MadGraph5\_aMC@NLO (v2.6.5) with the 
normalisation and factorisation scales set at $m_Z$ and shown in Fig:~\ref{fig:prod1}.
For further study, we choose 
the following benchmark points: $M_\Sigma=200$ GeV at $\sqrt s$ = 500 GeV, 
$M_\Sigma=300,400$ GeV at $\sqrt s$ = 1000 GeV, and 
$M_\Sigma=500, 600, 700$ GeV for $\sqrt s$ = 1500 GeV
\footnote {We did not go to masses beyond 700 GeV because it will require a 
3 TeV linear collider. At large energies, the analysis will require a detailed 
study of fatjets ($W/Z$) \cite{Das:2020gnt,Das:2020uer} which emerge as decay products of the
charged fermions.}. 

The decays of the singly charged fermions lead to the following final states involving W/Z bosons, 
%\bea
%\Sigma^+ &\rightarrow &\phi^+ \nu ~~~\rightarrow ~W^+ Z \nu ;  ~~~~~\Sigma^- \rightarrow ~\phi^- \nu ~~~\rightarrow ~W^- Z \bar{\nu} \n \\
%\Sigma^+ &\rightarrow &\phi^{++} l^- \rightarrow ~W^+ W^+ l^- ;  ~\Sigma^- \rightarrow ~\phi^- \nu ~~~\rightarrow ~W^- Z \bar{\nu} \n \\
%\Sigma^+ &\rightarrow &\phi^{++} l^- \rightarrow ~W^+ W^+ l^- ;  ~\Sigma^- \rightarrow ~\phi^{--} l^+ \rightarrow ~W^- W^- l^+ 
%\label{eqn:singdecay}
%\eea
%=======================
%========================

governed by the equations in Sec:~\ref{sec:model}. Among all final states, 
the leptonic final states or final states of 
leptons+jets suffer from lower effective cross-section due to small 
branching ratio of $W/Z$ into leptons. 
The signals with multiple jets 
have the advantage over multilepton states as the branching ratio 
of $W/Z$ is more into jets than leptons.
The final states involving multiple jets can have a maximum of 6 jets, 
coming from the decays of $W/Z$. Here, we show a detailed
analysis of two final states:
\vspace{-0.3cm}
\begin{itemize}
\item Channel(A): One lepton $(\ell^\pm)$ + 4 jets.
\item Channel(B): Two opposite sign lepton pair $(\ell^+ \ell^-)$ + 4 jets.
\end{itemize}
\vspace{-0.3cm}
These type of signals 
in $e^+e^-$ collider have a great chance for discovery due to smaller background, 
which is also shown in \cite{Azuelos:1993qu}.
%M1= singly charged VLL
%M2= doubly charged VLL
%M3= scalar masses (same for singly and doubly charged scalars)

\vspace{-0.3cm}

%==============================
\subsection{Backgrounds}
%--------------------------------

\vspace{-0.3cm}

The major backgrounds for the channels under study 
get contribution from di-boson ($WW$, $ZZ$), $t\bar t$, 
$t\bar V$, Triboson ($VVV$=$ZZZ$, $ZWW$) and $HZ$ production. 
The variation of these 
major backgrounds with $\sqrt{s}$ is already shown 
in \cite{Das:2020gnt}. Along with multileptons, as 
the channels under investigation  
include multiple jets, we demand {\it inclusive} cross-section 
of these backgrounds by producing at least two 
jets in association, such as, di-boson+ 2 jets production($VVjj$), 
$t\bar t$+ 2 jets and $HZ$+2 jets. The contribution from the 
$\ell\ell$+ 2 jets, 4-jets and 4-top production are 
found to be small. We include these backgrounds in ``others'' category. 
Among all the backgrounds, the cross-section of $ZZjj$ is found to be larger.

\vspace{-0.3cm}

%==================================================================
\subsection{Collider Analysis}
%===================================================================

\vspace{-0.3cm}

In order to generate events,  
we use MadGraph5\_aMC@NLO (v2.2.1)~\cite{Alwall:2014hca}, where the showering 
and hadronization are done in a similar way as mentioned before 
in the LHC part. In FastJet, the jets are reconstructed with distance parameter
$R=0.4$ using anti-$K_t$ algorithm. In Delphes, we use the Delphes ILD card \cite{Behnke:2013lya} 
for detector simulation.
The signal and background events are required to pass 
through selections on different kinametic distributions, as given 
in Table~\ref{tab:4}. At first, we select events with basic cuts, $A1$. 
Later, while selecting the single lepton or the oppositely charged lepton 
pair, we make sure that it is well isolated from the jets coming from the decays 
of $W/Z$ by requiring a moderate isolation cut in $A2$. We have also imposed a cut 
on $M(\ell^+, \ell^-)$ in channel (B) to reduce the background further.
%------------------
\begin{table}[tpb]
\begin{small}
  \begin{center}
\begin{tabular}{|c|c|c|}
\hline
   Selections & Channel (A) & Channel (B) \\
\hline
\bf A0 & $N(\ell) \geq 1$+ $N(j)\geq 4$                     &  $N(\ell) \geq 2$ + $N(j)\geq 4$                    \\
\hline
\bf A1 & $p_T(l)> 10$ GeV                            & $p_T(l)> 10$ GeV            \\
                 & $|\eta|(l) < 2.5$                &  $|\eta|(l) < 2.5$            \\
                 & $\Delta R(\ell, \ell/j) > 0.4$    & $\Delta R(\ell, \ell)> 0.4$            \\
                 &  $p_T(j)> 20$ GeV                                  & $p_T(j)> 20$ GeV            \\
                 &    $|\eta|(j) < 5.0$                                 &  $|\eta|(j) < 5.0$          \\
                 &   $\Delta R_{jj} > 0.4$                         &  $\Delta R_{jj} > 0.4$       \\
\hline        
\bf A2           &     $\Delta R(\ell,j) > 1.5$       & $\Delta R(\ell,j) > 1.5$ \\
                 &      --                                                  &  $M (\ell^+, \ell^-) > 100\gev$\\
\hline

\end{tabular}
\end{center}
\caption{\label{tab:4}Selections A1 and A2 for channel (A) and channel (B).}
\end{small}
\end{table}
%------------------------------------------------

\begin{figure}[tb]
\begin{center}
\includegraphics[width=6.0cm,height=5.7cm]{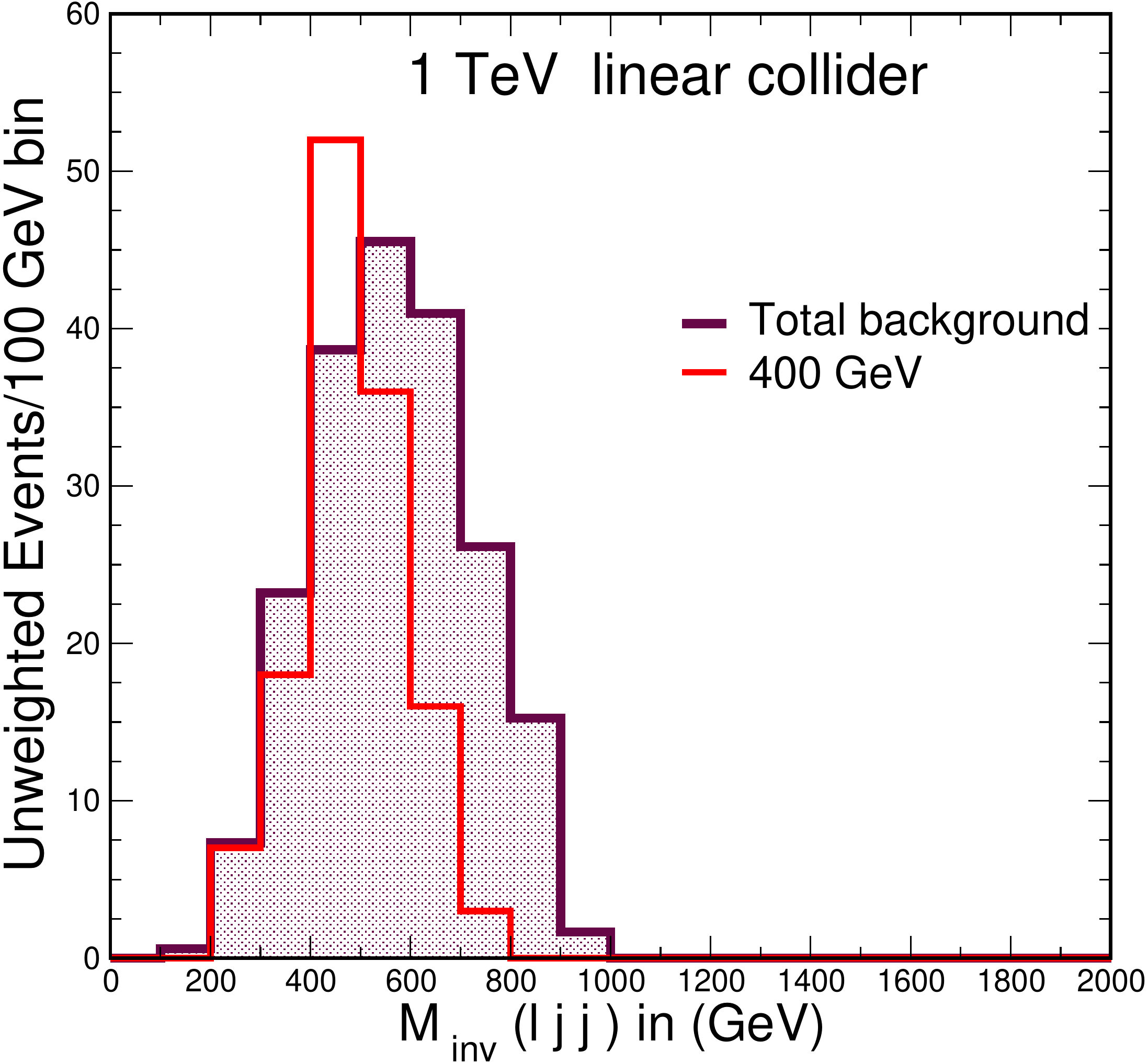}
\includegraphics[width=6.0cm,height=5.7cm]{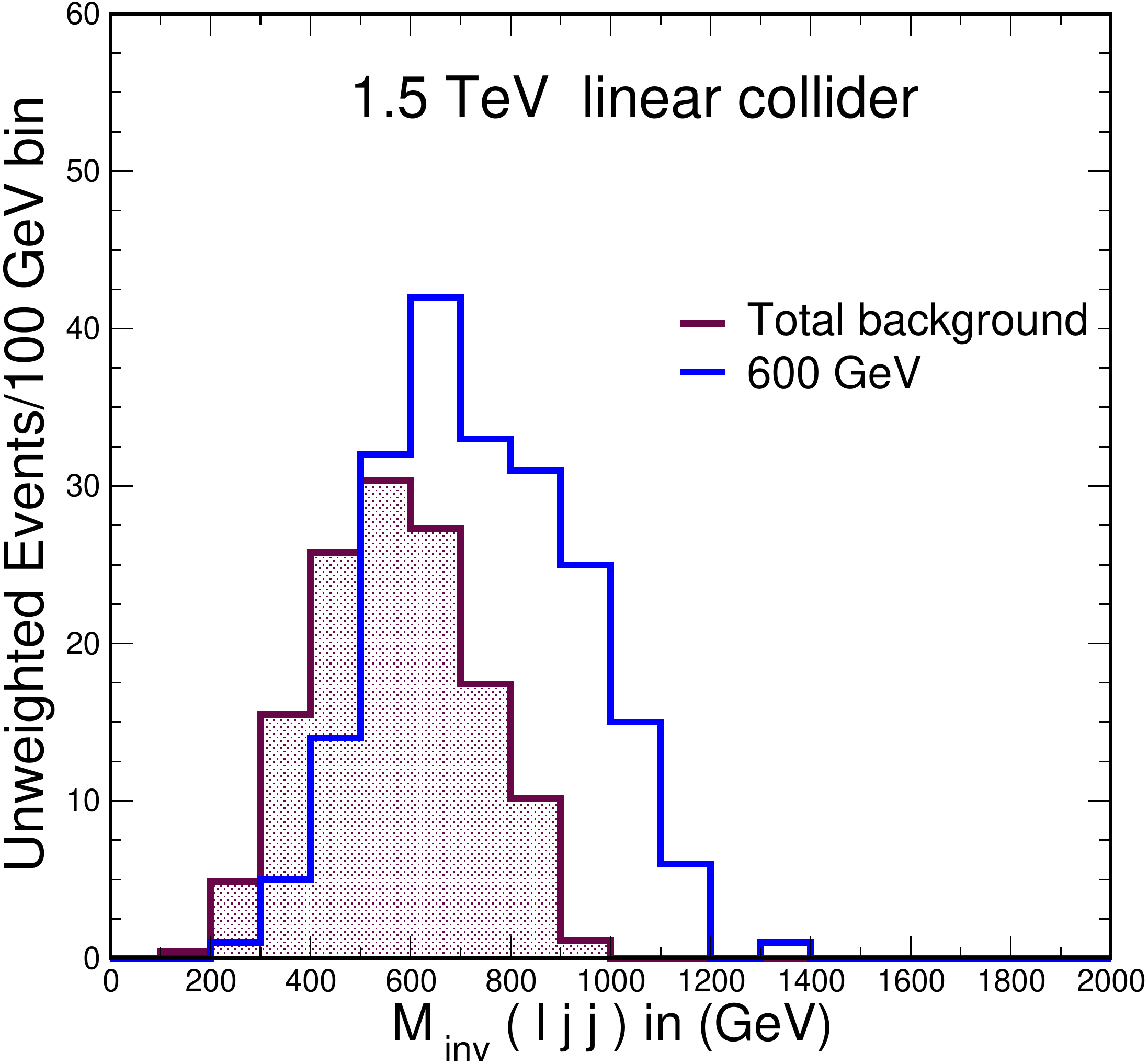}
\caption{Three body invariant mass 
$M(\ell j j)$ for $M_{\Sigma}=$ 400 GeV (left) and $M_{\Sigma}=$ 600 GeV (right) for channel (B) at 1 TeV and 1.5 TeV $e^+e^-$ collider 
respectively. The shadowed region represents the total background.}
\label{fig:dist}
\end{center}
\end{figure}

The signal and background cross-sections after the cuts are shown in 
Table~\ref{tab:5} and Table~\ref{tab:6}, respectively. 
We found the background to be small enough to give a very good signal to background ratio($S/B$), 
after the initial cuts $A1$ for channel (A). For channel (B), in order to improve the $S/B$ ratio,
we have imposed further cuts in $A2$ on selected opposite sign lepton pair ($\ell^+\ell^-$), as shown in Table~\ref{tab:4}.
The requirement of exactly two leptons with opposite sign in channel (B) makes the cross 
section smaller than channel (A). 
The largest background 
contribution comes from $t\bar{t} + jets$ due to the large cross-section. 
We further check that the additional cuts on kinematic variables such as $H_T$, $S_T$ or MET 
would reduce the signal efficiency effectively, hence we did not impose them.
%==================================================================
\begin{table}[htb!]
\centering
\begin{tabular}{|c|c|c|c|c|}
\hline
\hline
%\multirow{2}{5cm}{\multicolumn{3}{c|}{Benchmark Point} & {\bf Without Polarization} & \multicolumn{3}{c|}{\bf With Polarization}} \\
%~ & ~ & ~ & Without polarization &  \multicolumn{3}{c|}{\bf With Polarization}} \\
%\hline
%\cline{1-7}
\bf $\sqrt{s}$ & \bf $M_{\Sigma}$ (GeV) & \bf $\sigma$ (pb) & \bf $\sigma^{A}_{A2}$ (fb) & \bf $\sigma^{B}_{A2}$ (fb)\\
%\bf (GeV) & \bf (GeV) & \bf (GeV) & \bf $~$ & \bf $~$ & \bf (pb) \\
\hline
500 GeV & 200 & 0.706 & 4.45 & 0.049\\
\hline
& 200 & 0.218 & 15.70 & 0.131\\
1 TeV & 300 & 0.209  & 14.63 & 0.125\\
& 400 & 0.175 & 13.50 & 0.122\\
\hline
 & 500 & 0.089 & 7.56 & 0.107\\
1.5 TeV & 600 & 0.077 & 7.24 & 0.1001\\
 & 700 & 0.05 & 4.95 & 0.09\\
\hline
\hline
\end{tabular}
\caption{Cross-sections for the signal $e^+ e^- \rightarrow \Sigma^+ \Sigma^-$ before and after the selections. 
Channel (A) corresponds to $l^{\pm}$+ 4 jets and Channel (B) corresponds to $l^{\pm}l^{\pm}$+ 4 jets. 
%Polarizations are considered are (-0.3,0.8) for $\sqrt{s}$=500 GeV, (-0.2,0.8) for $\sqrt{s}$=1 TeV and (0.0,0.8) for $\sqrt{s}$=1.5 TeV.
}
\label{tab:5}
\end{table}
%==================================================================
\begin{table}[htb!]
\centering
\begin{tabular}{|c| rrr | rrr|}
\hline
\hline
%\multirow{2}{5cm}{\multicolumn{3}{c|}{Benchmark Point} & {\bf Without Polarization} & \multicolumn{3}{c|}{\bf With Polarization}} \\
%~ & ~ & ~ & Without polarization &  \multicolumn{3}{c|}{\bf With Polarization}} \\
%\hline
%\cline{1-7}
~ & \multicolumn{3}{c}{$\sigma^{A}_{A2}$ (fb)}  & \multicolumn{3}{c}{$\sigma^{B}_{A2}$ (fb)} \\ 
\hline
\bf Background & $\sqrt{s}=500$ GeV & 1 TeV & 1.5 TeV &  $\sqrt{s}=500$ GeV & 1 TeV & 1.5 TeV \\
\hline
$diboson+jets$ & 8.08 & 3.04 & 1.63 & 0.0 & 0.0 & 0.0\\
$t\bar{t} + jets$ & 82.5 & 24.75 & 11.25 & 1.1 & 0.33 & 0.15\\
$t\bar{t}V$ & 1.121 & 1.79 & 1.083 & 0.039 & 0.0614 & 0.037\\
$VVV$ & 2.85 & 5.0 & 3.67 & 0.0024 & 0.0035 & 0.0031\\
$HV + jets$ & 2.85 & 0.65 & 0.3 & 0.045 & 0.0 & 0.0\\
$others$ & -- & -- & -- & 0.045 & 0.0425 & 0.035\\
\hline
\hline
$Total$ & 97.4 & 34.35 & 17.94 & 1.186 & 0.437 & 0.225\\
\hline
\end{tabular}
\caption{Cross-sections for various backgrounds corresponding 
to channels A ($l^{\pm}+ 4 jets$) and B ($l^{\pm}l^{\pm}+ 4 jets$) after the selections.}
\label{tab:6}
\end{table}
%==================================================================
Even though the signal cross section in channel (B) is less, we find that it is an excellent channel to 
reconstruct the invariant mass of $M_\Sigma$ from the $\ell^+ \ell^- j j$ distribution. 
We show the distribution for two cases, $M_{\Sigma}$=400 and 600 GeV at $\sqrt{s}=$1 and 1.5 TeV
respectively in Fig:\ref{fig:dist}.
%==================================================================
%=================
%{\color{red} How jets are handled at linear collider.}\\
% {\color{red} Detail about the detector card.}

\begin{figure}[tb]
\begin{center}
\includegraphics[width=6.0cm,height=5cm]{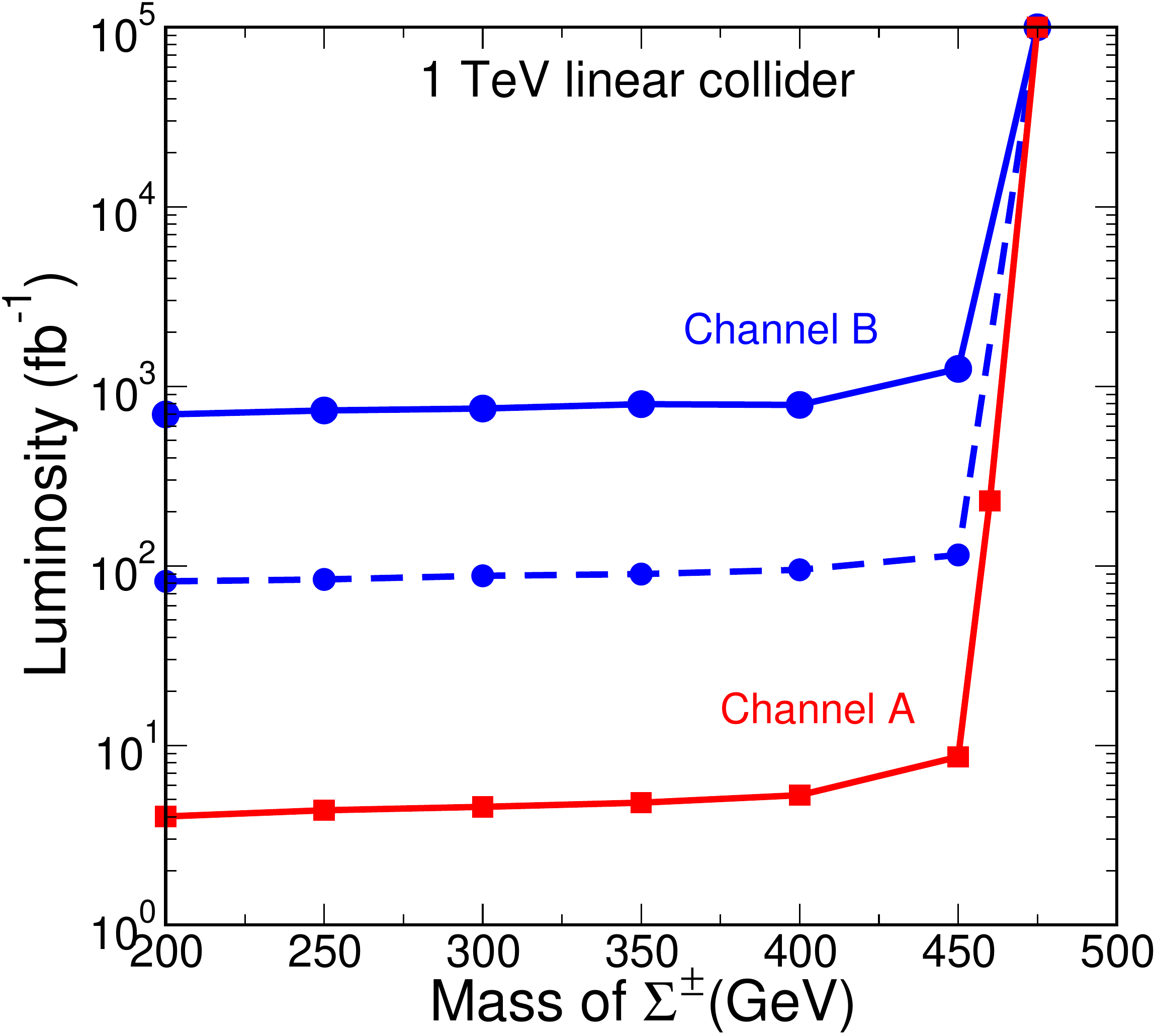}
\includegraphics[width=6.0cm,height=5cm]{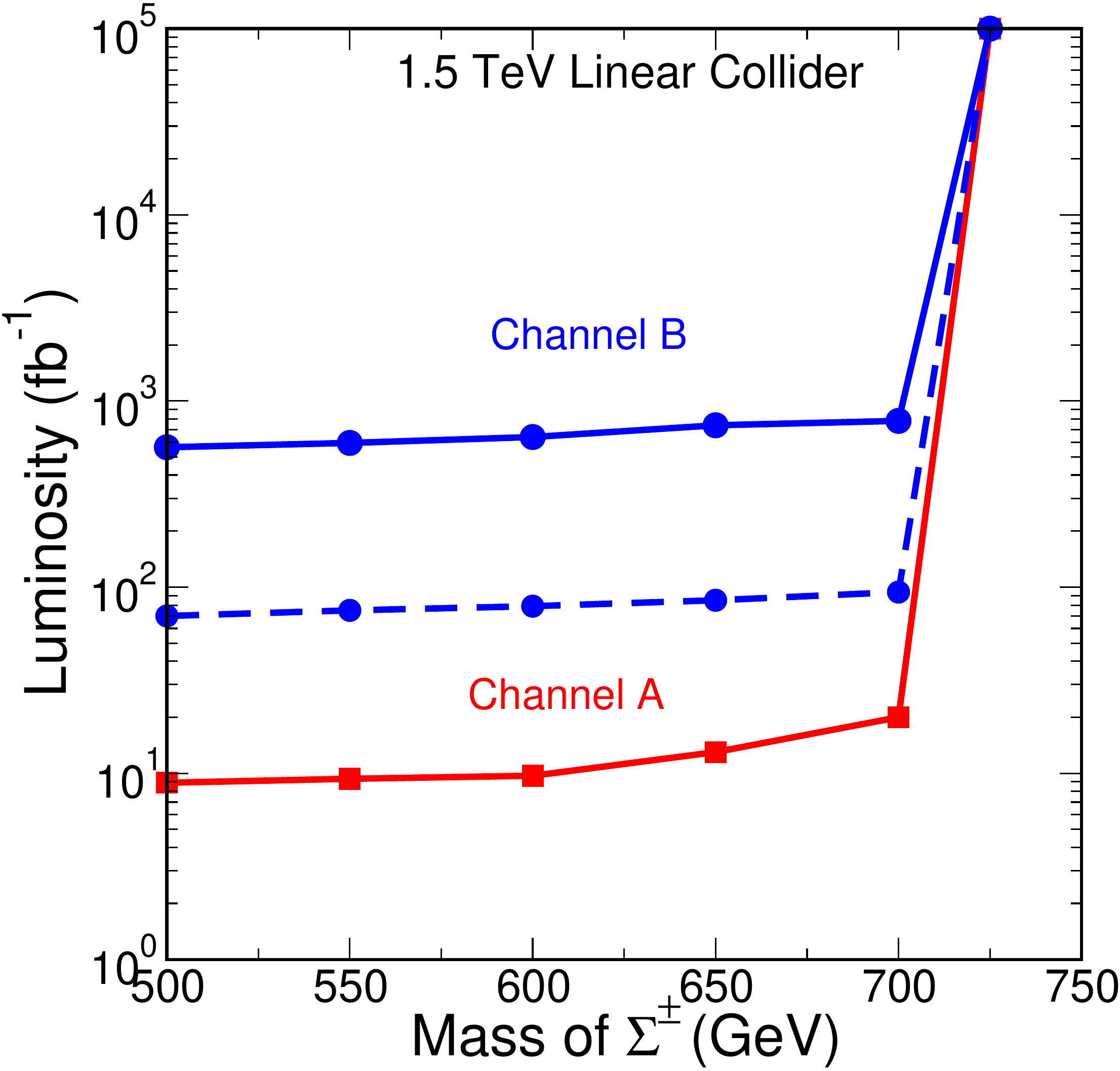}
\caption{The integrated luminosity for 
discovery (Solid line) as a function 
of the mass of the singly charged fermion ($M_\Sigma$) is shown 
for channel (A) and (B) at 1 TeV (left) and 1.5 TeV (right) $e^+e^-$ collider. 
The integrated luminosity for exclusion (dotted line) is also shown for channel B.}
\label{fig:result2}
\end{center}
\end{figure}

\vspace{-0.3cm}
%==============================
\paragraph{\underline {Result}}
%==============================
For the study in $e^+e^-$ collider, the background is larger 
compared to the LHC scenario. Hence, Eq:\ref{eq:zdis1} reduces 
to a simple form of $S/\sqrt{B}$. The integrated luminosity for 
discovery as a function 
of the mass of the singly charged fermion is shown in 
Fig:\ref{fig:result2}. We find that, 
the discovery potential of channel (A) is much better than channel (B), i.e,
the required integrated luminosity is less in channel (A),
for $M_\Sigma\leq 450$ GeV and $M_\Sigma\leq 750$ GeV, 
at $\sqrt{s}=1$TeV and 1.5 TeV respectively. 
With $\leq 20$ fb$^{-1}$ luminosity, it is possible to discover
in the region of $M_\Sigma \leq 700$ GeV with $5\sigma$ significance
in channel (A). Whereas, the required luminosity for $5\sigma$ 
discovery is $\leq 1000$ fb$^{-1}$, for the same in channel (B).
For 95\% exclusion limit, the entire mass region can be probed with luminosity less than 
100 fb$^{-1}$ in channel (B). The required luminosity in channel (A) is very small for the same, 
hence we do not plot them. 
%==============
%=================

%=====================================================================
\section{Conclusion and Outlook}
\label{sec:outlook}
%======================================================================
We have discussed the discovery potential of the singly and the doubly
charged fermions, which are components of a quintuplet, at the LHC and future 
$e^+e^-$ colliders. Such a specific model, as we have considered, with quintuplet fermions 
and a scalar multiplet, predicts certain signatures which require alternate search 
strategies.

In the study of signatures at the LHC, we have discussed the possible multilepton 
and multi(lepton+jet) signatures of the doubly charged fermions, as they have larger 
cross-sections compared to that of the singly charged fermions. 
For the doubly charged quintuplet fermion ($\Sigma^{\pm\pm}$), 
5$\sigma$ discovery might be possible at integrated luminosity of 
3000 fb$^{-1}$ at 14 TeV LHC if $M_\Sigma\leq 980$ GeV. The exclusion 
limit can be extended upto 1.2 TeV with the same parameters. 

\begin{figure}[tb]
\begin{center}
\includegraphics[width=5.3cm,height=5cm]{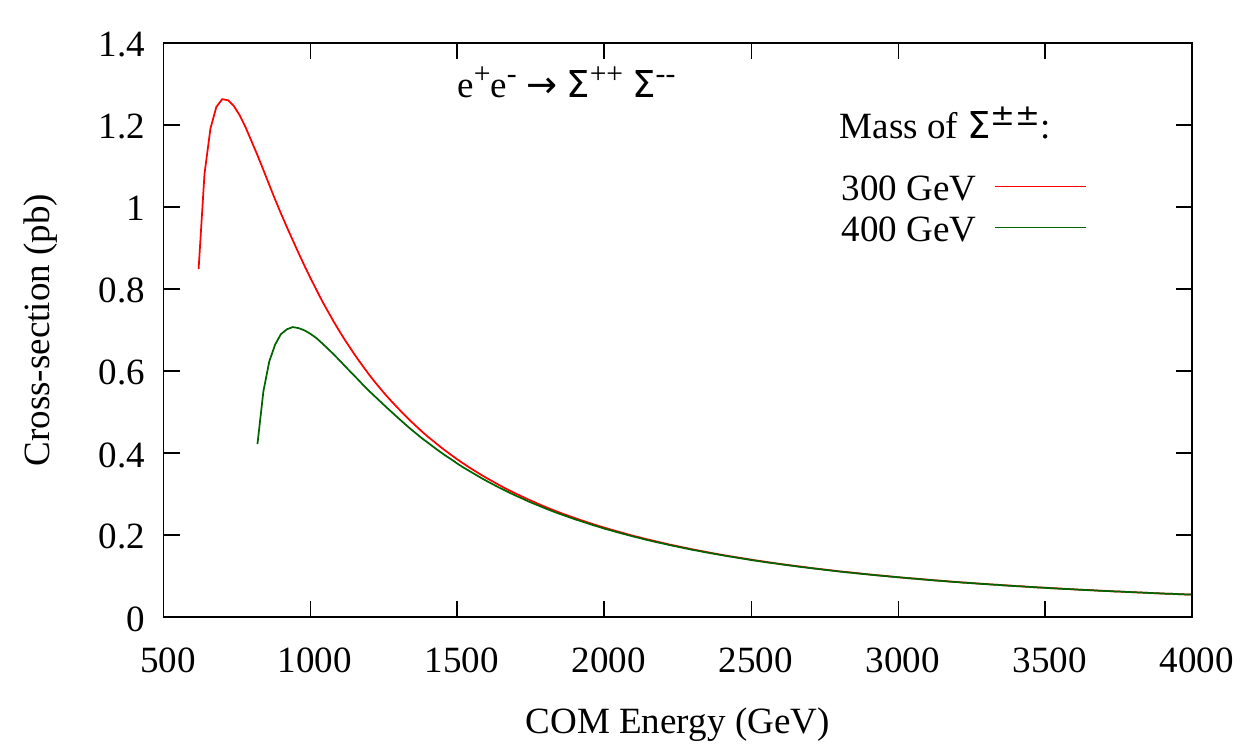}
\includegraphics[width=5.3cm,height=5cm]{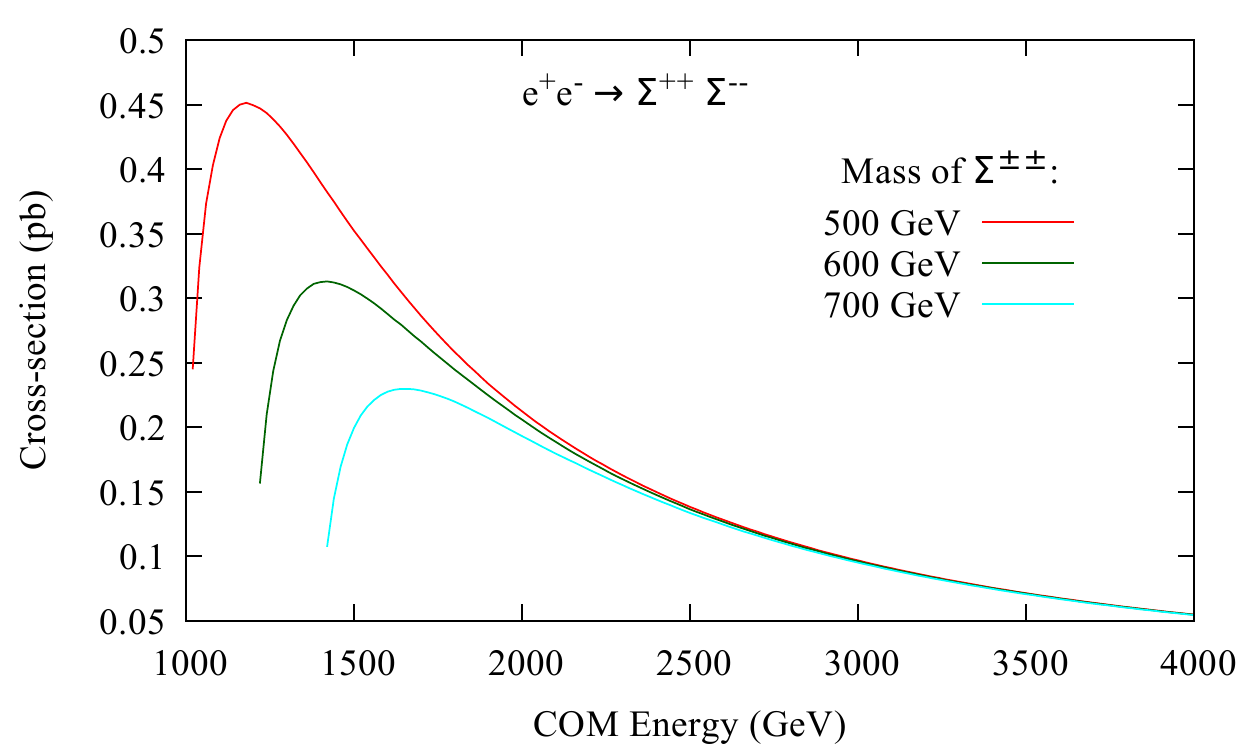}
\includegraphics[width=5.3cm,height=5cm]{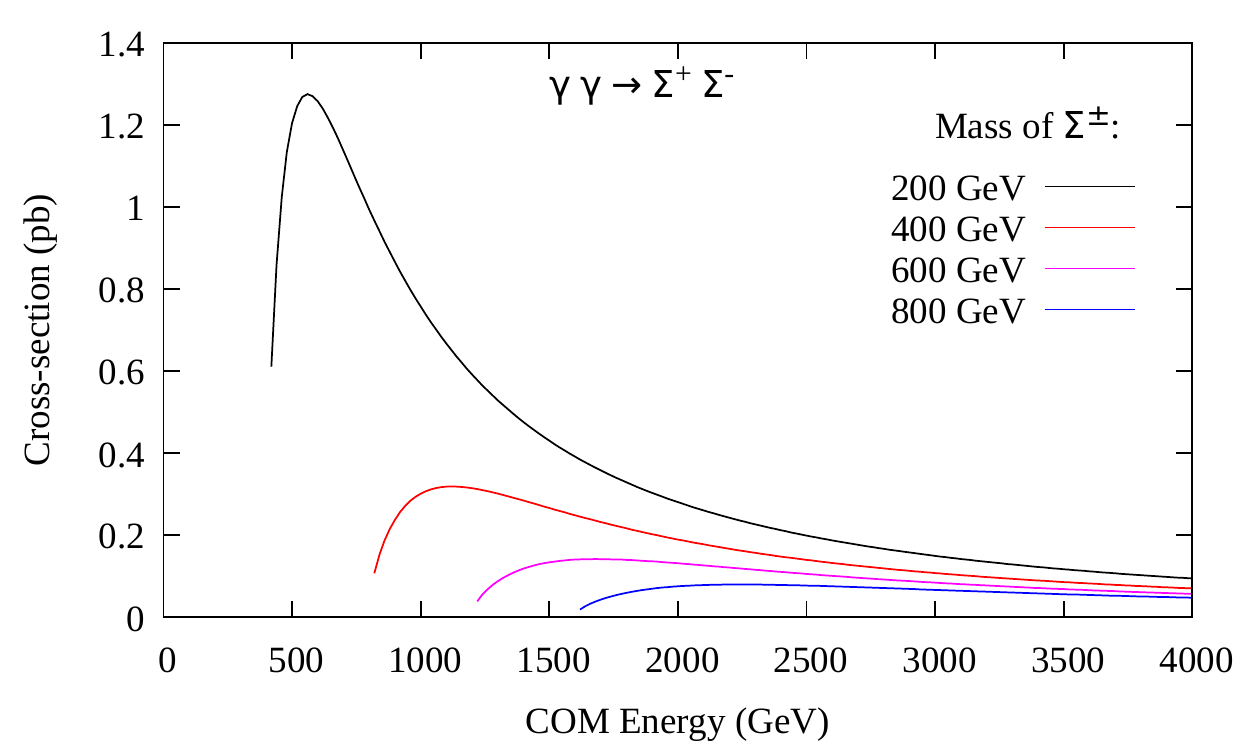}
\caption{(Left and Middle) The doubly-charged quintuplet fermion production cross-section as a function of the center of mass energy at $e^+e^-$ collider. (Right) Cross-section for the singly-charged quintuplet fermion production at $\gamma \gamma$ collider. }
\label{fig:eess2}
\end{center}
\end{figure}

On the other hand, linear colliders, such as the $e^+e^-$ collider, offer 
a much cleaner environment to study the signatures associated with multiple jets. Thus,
the signals have the advantage of a larger cross-section $\times$ BR, where the $W/Z$ bosons decay 
into jets. We find that the singly charged fermion ($\Sigma^\pm$) shows a great discovery potential 
at the $e^+e^-$ collider, unlike the case of the LHC.
There might be a possibility of 5 $\sigma$ discovery with 1000 fb$^{-1}$ 
luminosity at $e^+e^-$ collider for $M_\Sigma\leq 700$ GeV. 
Similar kind of final states also exist for the doubly charged quintuplet fermion 
but with more leptons and jets, making the analysis much more complicated. 
Thus, we will address it somewhere else.
%The doubly charged fermions also produce similar kind 
%of final states but with more leptons and jets, making the analysis much more complicated, which 
%we will address somewhere else. 
The cross-sections for the pair production of the doubly charged fermions 
at the $e^+e^-$ collider are shown in
Fig:~\ref{fig:eess2} (left and middle).
%=======================
%========================  

An $e^+e^-$ linear collider can also be operated as a $\gamma \gamma$ and 
an $e^- \gamma$ collider, as illustrated in Ref:\cite{osti_378863,DeRoeck:2003cjp}. 
The highly intense photons for the collision are obtained by Compton back-scattering 
laser photons on intense high-energy electron beams. 
Due to the coupling with photon, charged particles can be produced with a considerably high cross-section at 
these photon colliders. In the present model, the production of the 
singly charged quintuplet fermions is possible via 
$\gamma \gamma \rightarrow \Sigma^{+} \Sigma^{-}$,
$e^- \gamma \rightarrow \Sigma^{+} \phi{--}$ and 
$e^- \gamma \rightarrow \Sigma^{-} \phi^{0^*}$ modes, along with the conjugate process in each case.
The production cross-section for the singly-charged quintuplet fermion at 
$\gamma \gamma$ collider is shown in Fig:~\ref{fig:eess2} (right) as a function 
of the center of mass energy. 
These production modes, alone or combined with 
$e^+e^-$ collision, show a great potential at future linear colliders. Over all, 
the nonstandard decay modes of the quintuplet fermions offer different signals which 
require alternate search strategies and there might be an opportunity for discovery and/or 
exclusion at the HL-LHC and future linear colliders. 

\vspace{-0.3cm}
%---------------------------------------------
\section*{Acknowledgments}
\vspace{-0.5cm}
%\vspace{0.5cm} 
N.K. acknowledges the support from the Dr. D. S. Kothari Postdoctoral scheme (201819-PH/18-19/0013). V.S. thanks Brajesh Choudhary for access to computers bought under the aegis of the Grant No. SR-MF/PS-0212014-DUB (G) of the DST (India).

\vspace{-0.3cm}
%=============================
\bibliographystyle{unsrt}
\bibliography{ref}
%=============================

\end{document}